\documentclass[showpacs,preprint]{revtex4}
\usepackage{graphicx}
\usepackage{dcolumn}
\usepackage{amsmath}
\usepackage[latin1]{inputenc}
\usepackage{graphicx, psfrag}
\usepackage{amssymb}
\usepackage[colorlinks=true, citecolor=blue, urlcolor = blue, linkcolor= red, bookmarks=true]{hyperref}
\usepackage{float}
\usepackage{amsmath}
\usepackage{amsfonts}
\usepackage{dcolumn}
\usepackage{hyperref}
\usepackage{subfigure}
\usepackage{epstopdf}
\usepackage{booktabs}
\usepackage{multirow}

\makeatletter
\def\btt#1{\texttt{\@backslashchar#1}}
\DeclareRobustCommand\bblash{\btt{\@backslashchar}} \makeatother

\makeatletter
\def\btt#1{\texttt{\@backslashchar#1}}
\DeclareRobustCommand\bblash{\btt{\@backslashchar}} \makeatother

\begin{document}
\title{Shadows of quintessential dark energy black holes in the domain of outer communication }
\author{Balendra Pratap Singh$^{a}$}\email{balendra29@gmail.com}

\affiliation{$^{a}$Department of Applied Sciences and Engineering,  Tula's Institute, Dehradun, Uttarakhand 248197, India}

\begin{abstract}
The rotating black holes in the quintessential dark energy correspond to three horizons: inner, outer, and quintessential horizon. The domain of outer communication is the region between  outer and quintessential horizon. Here, in this work we study the photon region and shadows of the quintessential dark energy black holes when the observer stays statically in the domain of outer communication. The quintessential dark energy black holes shadow characterizes by its mass $(M)$, spin parameter $(a)$, quintessential dark energy parameter $(\omega_q)$, and normalization factor $(\gamma)$. The dark energy parameter $\omega_q$ can take values in between  $-1.1<\omega_q<-1/3$ and follows the equation of state $\omega_q$=pressure$(p)$/energy density($\rho_q)$. This state parameter significantly affects the shape and size of the black hole shadow. We generalize all the geodesic equations of motion for $\omega_q$ and obtain relation to visualize the black hole shadow by a static observer at any arbitrary distance in the domain of outer communication. We analytically estimate the black hole shadow observables: radius $R_s$, distortion parameter $\delta_s$ and the shadow area $A$. Using the numerical values of shadow radius $R_s$ and area $A$, we  obtain the angular diameter of the black hole shadow. The angular size  of the M87   and Sgr A$^*$ black holes are $\ 42 \pm 3 \mu a s$ and $48.7 \pm 7  \mu a s $ respectively as observe by Event Horizon Telescope (EHT). In this case, the angular diameter of the black hole shadow   increases with the quintessence parameter $\omega_q$ and takes values $\theta_d \approx 20 \pm 3{^o}$ with the parameter $-0.66 \leq \omega_q \leq -0.62$ for the static observer at $r_o=5M$ in the domain of outer communication. 
 
\end{abstract}
\maketitle
 {\section{Introduction}}
The observations of type Ia supernovae (SNe Ia) lie between the red shift range $0.16\leq z \leq 0.62$, confirming that our Universe is going under the late time acceleration \cite{SupernovaSearchTeam:1998fmf,SupernovaCosmologyProject:1998vns}.  According to general relativity, this cosmic acceleration indicates that there exists some strange energy component in the Universe which is called dark energy. The observational constrains over the state parameter $\omega_q$ provided by the large state structure of the Universe and the cosmic microwave background (CMB) is $-1.1<\omega_q<-1/3$ \cite{suzuki:2012, Planck:2015bue, Chiba:2012cb}.  The hypothesis of dark energy is compatible with the standard model of big bang cosmology ($\Lambda$CDM model) when the dark energy state parameter  is exactly equal to -1. For this value of state parameter, the dark energy is considered as cosmological constant, which is interestingly agreements with the observations but still there is a possibility that some significant component of the dark energy densities have state parameters other than -1 \cite{Caldwell:1997ii}. One of the simplest is the quintessence dark energy model in which the dynamical scalar field is minimally  coupled with the gravity \cite{Caldwell:1997ii}. The quintessence dark energy is dynamic and time-varying which is  different from the cosmological constant model which does not change with time. Some researchers also consider the quintessence dark energy as the fifth fundamental force responsible for the expansion of the Universe \cite{Carroll:1998zi,Cicoli:2012tz, Dvali:2001dd}. 

The asymptotic structure of the black hole gets modified in the presence of the quintessential dark energy. The black hole spacetime remains no more asymptotically flat in quintessence due to the cosmological horizon. The very first model of the  black hole in quintessence was presented by Kislev \cite{Kiselev:2002dx}. After that, several researchers intensively studied the properties of spherically symmetric black holes in quintessence dark energy  \cite{ChenWang:2008, Ding:2013, Fernando:2012, Wang:2009, Jamil:2015, Kalam:2014, Xiao:2010, Ghaderi:2015a, Ghaderi:2015b, Biswas:2015, Thomas:2012, Uniyal:2015, Kuriakose:2009, Kar:2006, Pandey:2014ona, Priyabrat:2015, Singh:2014a, Singh:2014b, Xu:2016ylr, Xu:2017vse}. The Lovelock black holes in quintessence have been studied by \cite{Ghosh:2017cuq, Toledo:2019}. The study of Narnia black holes in quintessence has been done by \cite{Fernando:2013mex}. The authors of \cite{Pedraza:2020uuy} discussed the geodesics of the Hayward black hole in quintessence. Thermodynamics of the Bardeen black hole in quintessential dark energy studied in \cite{Wu:2022fkz}. The rotating counterpart of the spherically symmetric black hole in quintessence obtained by \cite{Toshmatov:2015npp} and \cite{Ghosh:2015ovj}.  The study of rotating anti-de-sitter and rotating charged anti-de-sitter  black holes in the presence of perfect fluid matter have been intensively studied by \cite{Xu:2017bpz} and \cite{Xu:2017vse}. The author of \cite{Belhaj:2020rdb} extended Schwarzschild black holes in quintessence up to  $D$-dimensional spacetimes and studied gravitational lensing and shadow properties.

The observational results from the Event Horizon Telescope (EHT)  proved that black holes are just not only  theoretical concepts. The first picture of the Messier 87 (M87) black hole revealed by the EHT group in 2019 \cite{Akiyama:2019cqa,Akiyama:2019brx,Akiyama:2019sww,Akiyama:2019bqs,Akiyama:2019fyp,Akiyama:2019eap}. Black holes are completely dark objects but interestingly they cast a shadow \cite{Goddi:2016qax}. The incoming photons towards the  black hole horizon which have quite large angular momentum fall inside the black hole event horizon and create a dark spot, and photons that carries a little bit smaller angular momentum form the photon region around the black hole \cite{Goddi:2016qax}. This dark spot surrounded with the bright photon rings is called a black hole shadow \cite{Perlick:2021aok}. Casting shadow by the black hole  also verifies the existence of the event horizon \cite{Akiyama:2019cqa}. Recently EHT collaborators published the image of Sgr A$^{*}$ black hole shadow. This supermassive black hole lies at the heart of our galaxy Milky way \cite{Akiyama:2022L12,Akiyama:2022L13,Akiyama:2022L14,Akiyama:2022L15,Akiyama:2022L16,Akiyama:2022L17}. The first analytical study of this subject was done by Synge \cite{Synge:1966} and Luminet \cite{Luminet:1979}. Later Bardeen extended it for the rotating spacetime \cite{Bardeen1}.   
In the past few years, several researchers analytically studied the black hole shadow for rotating and non-rotating spacetimes.\cite{Amarilla:2010zq,Amarilla:2013sj,Yumoto:2012kz,Abdujabbarov:2016hnw,Amir:2016cen,Tsukamoto:2014tja,Bambi:2010hf,Takahashi:2005hy,Wei:2013kza,Abdujabbarov:2012bn,Amarilla:2011fx,Bambi:2008jg,Atamurotov:2013sca,Wang:2017hjl,Schee:2008kz,Grenzebach:2014fha,Abdujabbarov:2015xqa,Cunha:2015yba,Cunha:2016bpi,Singh:2017xle, Singh:2022dqs,Kumar:2017tdw, Kumar:2019ohr}. The optical properties of the rotating and non-rotating black holes in the presence of the plasma medium have been studied by \cite{Perlick:2015vta, Atamurotov:2015nra,Abdujabbarov:2015pqp}
This subject has been extended for the higher dimensional spacetime by \cite{Papnoi:2014aaa,Abdujabbarov:2015rqa,Singh:2017vfr,Amir:2017slq, Belhaj:2022cdh}. Various researchers estimated black hole parameters using the properties of the black hole shadow \cite{KumarWalia:2022ddq,KumarWalia:2022aop,Vagnozzi:2022moj,Ghosh:2022jfi,Afrin:2021imp}.

The primary goal of this work is to analytically study the angular size of the  photon regions for rotating black holes in quintessential dark energy in the domain of outer communication. The rotating  black hole in quintessence corresponds to three horizons: inner, outer, and quintessential or cosmological event horizons \cite{Toshmatov:2015npp}. The domain  of outer communication is the region where the observer statically stays in between the outer and the cosmological horizon \cite{Perlick:2021aok}. The quintessence field affects the shadow region, and the appearance of the bright photons region varies as seen by the observer. The size of the black hole shadow decreases and gets distorted  with the state parameter $\omega_q$. The photon region will appear deviated and  more close to the event horizon in the presence of the quintessential dark energy. 

This paper is organized as follows: In Sec.~\ref{secone}, we discuss the rotating and non-rotating black hole metric with the quintessential dark energy. In Sec.~\ref{sectwo}, we derive the effective potential of the black hole and find the impact parameters. We analytically study the black hole shadow in Sec.~{\ref{secthree}} and for better understating of the effect of quintessential dark energy on the black hole shadow, we determine the shadow observables and angular size in Sec.~{\ref{secfour}}. Finally, we conclude our results in Sec.~\ref{secfive}.
 {\section{ Black Hole Metric}\label{secone}}
 {\subsection{Non-rotating black hole metric  in the presence of quintessential dark energy}}
\begin{figure*}
\begin{center}
    \begin{tabular}{c c}
	\includegraphics[scale=0.45]{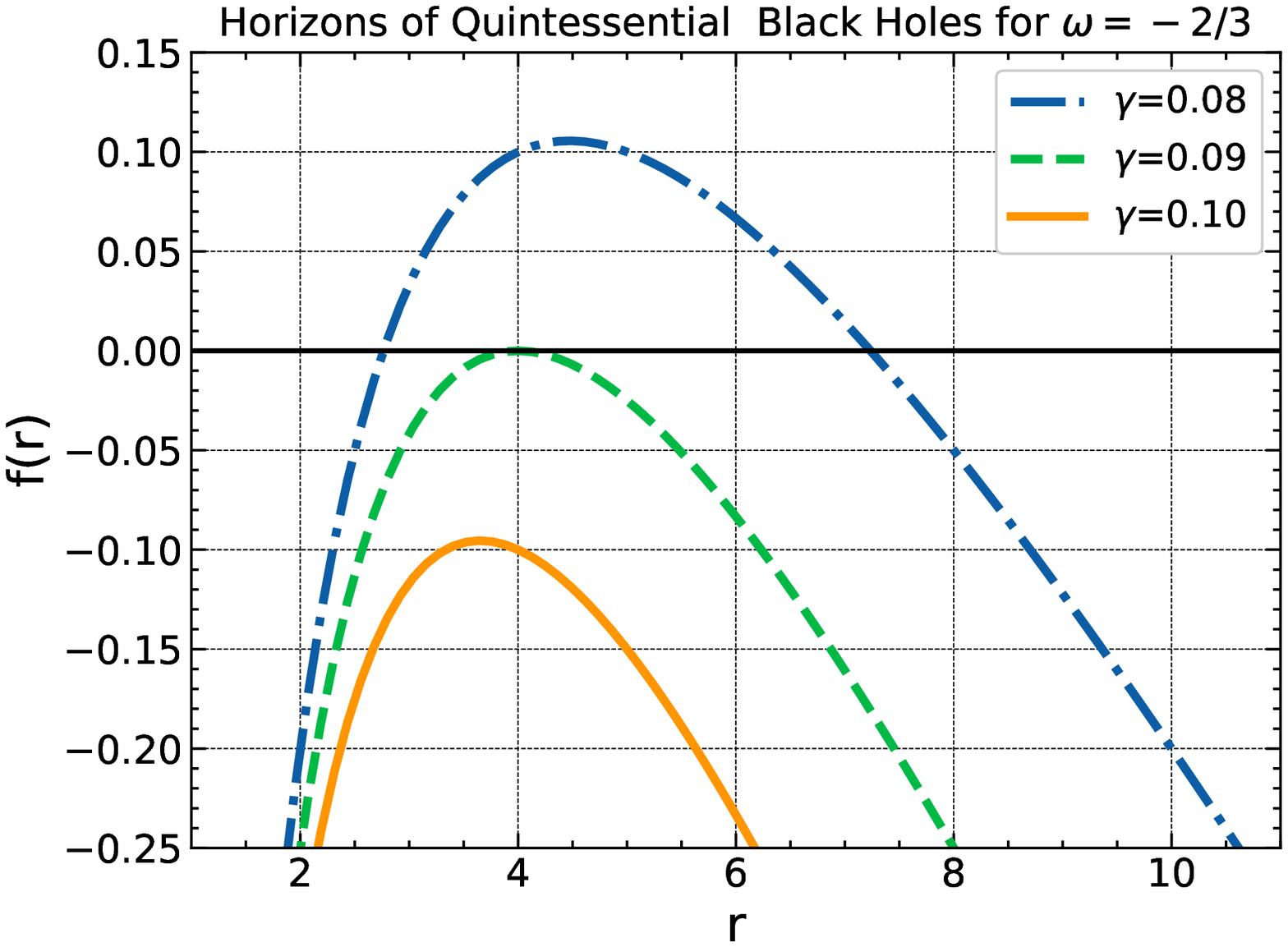} 
	\includegraphics[scale=0.45]{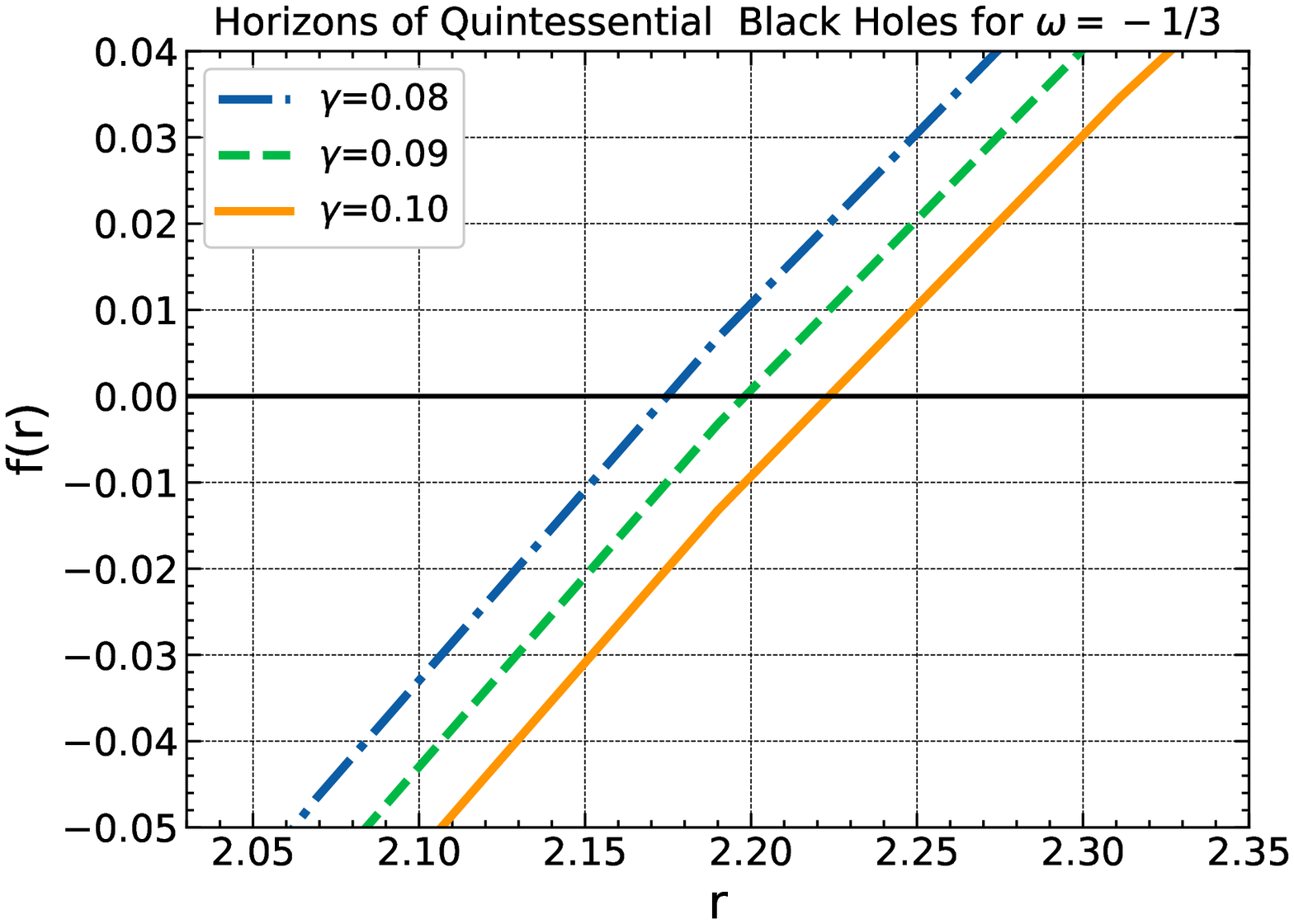}\\
	\includegraphics[scale=0.45]{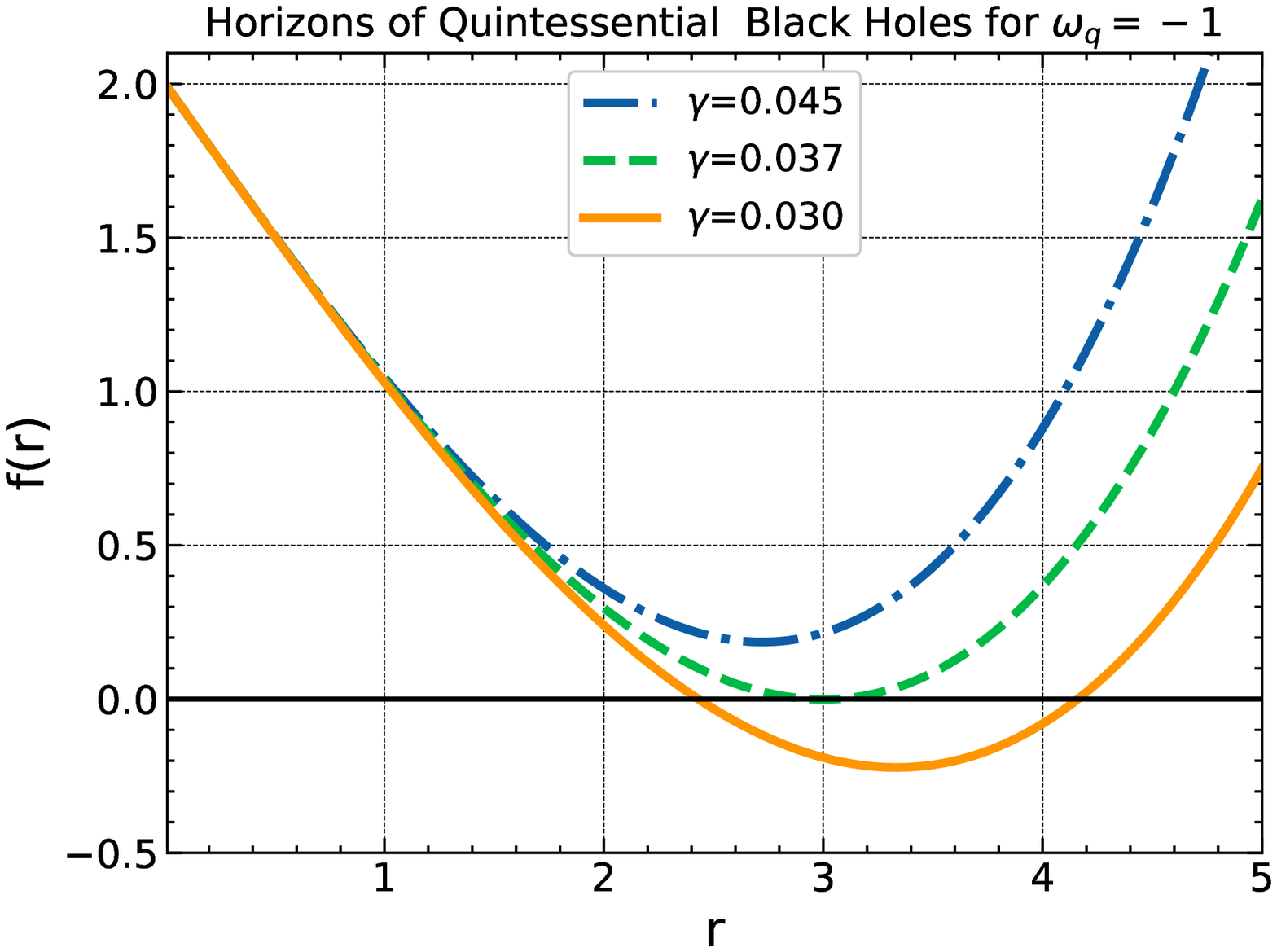}
	 \end{tabular}
    \caption{\label{fig1} Plot showing the horizons of the non-rotating black holes .}
    \end{center}
\end{figure*}
The spherically symmetric black hole metric with quintessential dark energy has been derived by Kislev \cite{Kiselev:2002dx}. In Boyer Lindquist coordinates and natural units $G=c=1$, the black hole metric is given by
\begin{eqnarray}\label{01}
ds^2=-f(r)dt^2+f^{-1}(r)dr^2+ r^2(d\theta^{2}+ \sin^{2}\theta d\phi^{2}),
\end{eqnarray}
with  
\begin{eqnarray}\label{02}
f(r)=1-\frac{2M}{r}-\frac{\gamma}{r^{3\omega_q+1}},
\end{eqnarray}
where $M$ is the black hole, $\gamma$ is the normalization factor and $\omega_q$ is the quintessential state parameter. In Fig.~{\ref{fig1}}, we plot the metric function $f(r)$ with radial distance $r$ for different values of $\omega_{q}$ and $\gamma$. For $\omega_q=-2/3$, the   quintessential dark energy significantly  affects the black hole horizons. Non-degenerate and degenerate horizons exist for the corresponding values of normalization parameter $\gamma$.  If we consider $\omega_q=-1/3$, the black hole metric reduces for the Kottler spacetime also degenerate and non-degenerate horizons exist with the corresponding values of $\gamma$. For $\omega_q=-1$, the black hole metric corresponds single horizon (cf.~Table~\ref{table1} and Fig.~\ref{fig1}). In the absence of the quintessential field the black hole metric (\ref{01}) reduces for the Schwarzschild spacetime.
\begin{table}[]
\begin{tabular}{|ccc|ccc|cc|l}
\cline{1-8}
\multicolumn{3}{|c|}{\textbf{$\omega_{q}=-2/3$}} & \multicolumn{3}{c|}{\textbf{$\omega_{q}=-1/3$}} & \multicolumn{2}{c|}{\textbf{$\omega_{q}=-1$}} & \multicolumn{1}{c}{} \\ \hline\hline \cline{1-8}
\multicolumn{1}{|c|}{\textbf{$\gamma$}} & \multicolumn{1}{c|}{\textbf{Inner Horizon}} & \textbf{Outer Horizon} & \multicolumn{1}{c|}{\textbf{$\gamma$}} & \multicolumn{1}{c|}{\textbf{Inner Horizon}} & \textbf{Outer Horizon} & \multicolumn{1}{c|}{\textbf{$\gamma$}} & \textbf{Horizon} &  \\ \cline{1-8}
\multicolumn{1}{|c|}{0.030} & \multicolumn{1}{c|}{no horizon} & no horizon & \multicolumn{1}{c|}{0.100} & \multicolumn{1}{c|}{2.763932} & 7.236067 & \multicolumn{1}{c|}{0.080} & 2.173913 &  \\ \cline{1-8}
\multicolumn{1}{|c|}{0.037} & \multicolumn{1}{c|}{2.946526} & 3.056142 & \multicolumn{1}{c|}{0.120} & \multicolumn{1}{c|}{4.000000} & 4.000000 & \multicolumn{1}{c|}{0.900} & 2.197802 &  \\ \cline{1-8}
\multicolumn{1}{|c|}{0.045} & \multicolumn{1}{c|}{2.752125} & 2.752125 & \multicolumn{1}{c|}{0.150} & \multicolumn{1}{c|}{no horizon} & no horizon & \multicolumn{1}{c|}{0.100} & 2.222222 &  \\ \cline{1-8}
\end{tabular}
\caption{The horizons of non-rotating black holes in quintessential dark energy}
\label{table1}
\end{table}
 {\subsection{ Rotating black holes in quintessential dark energy}}
   \begin{figure}
\begin{center}
    \begin{tabular}{c c }
	\includegraphics[scale=0.7]{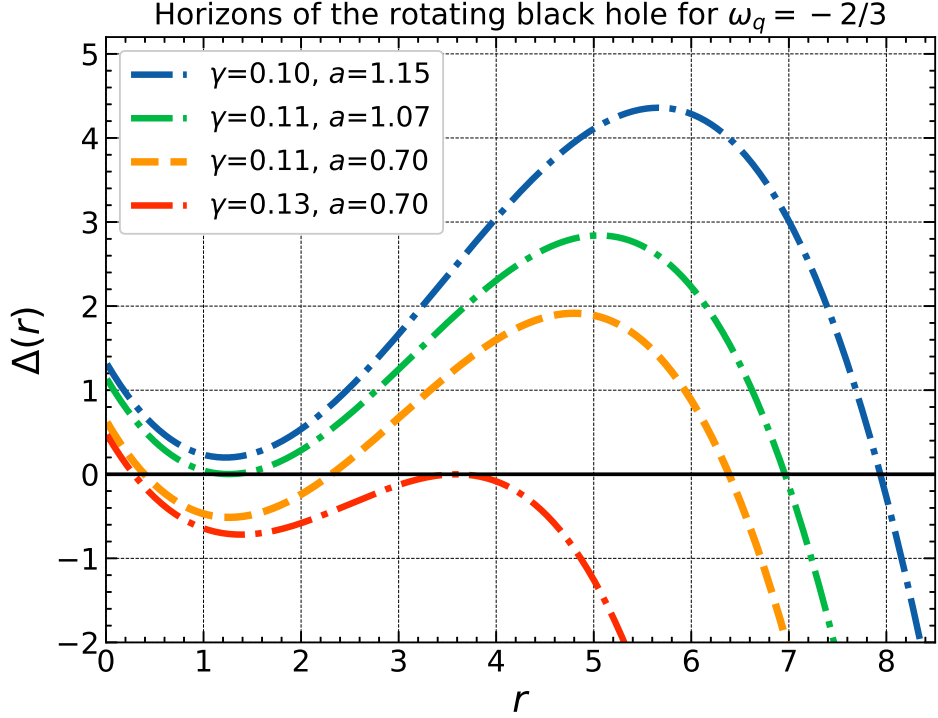} 
	 \end{tabular}
    \caption{\label{fig2} Plot showing the horizons of the rotating black hole in quintessential dark energy.}
    \end{center}
\end{figure}
\begin{table}[]
\begin{tabular}{|ccccc|}
\hline\hline
\multicolumn{5}{|c|}{$\omega_q=-2/3$} \\ \hline\hline
\multicolumn{1}{|c|}{$a$} & \multicolumn{1}{c|}{$\gamma$} & \multicolumn{1}{c|}{\begin{tabular}[c]{@{}c@{}}Inner \\ Horizon\end{tabular}} & \multicolumn{1}{c|}{\begin{tabular}[c]{@{}c@{}}Outer \\ Horizon\end{tabular}} & \begin{tabular}[c]{@{}c@{}}Quintessential\\ Horizon\end{tabular} \\ \hline
\multicolumn{1}{|c|}{0.70} & \multicolumn{1}{c|}{0.13} & \multicolumn{1}{c|}{0.283718} & \multicolumn{1}{c|}{3.590570} & 3.590570 \\ \hline
\multicolumn{1}{|c|}{0.70} & \multicolumn{1}{c|}{0.11} & \multicolumn{1}{c|}{0.394402} & \multicolumn{1}{c|}{2.309770} & 6.386736 \\ \hline
\multicolumn{1}{|c|}{1.07} & \multicolumn{1}{c|}{0.11} & \multicolumn{1}{c|}{1.263181} & \multicolumn{1}{c|}{1.263181} & 6.564545 \\ \hline
\multicolumn{1}{|c|}{1.15} & \multicolumn{1}{c|}{0.10} & \multicolumn{1}{c|}{No Horizon} & \multicolumn{1}{c|}{No Horizon} & 7.598965 \\ \hline
\end{tabular}
\caption{\label{table2}Table showing the numerical values for the horizons of the rotating black hole in quintessential dark energy}
\end{table}
The rotating black hole metric surrounded with the quintessential dark energy is
\begin{eqnarray}\label{03}
ds^2&=&-\left(1-\frac{2Mr+ \gamma r^{1-3\omega_q}}{\Sigma}\right)dt^2+\frac{\Sigma}{\Delta}dr^2-2a\sin^2\theta\left(\frac{2Mr+ \gamma r^{1-3\omega_q}}{\Sigma}\right)d\phi dt+\Sigma d\theta^2 \nonumber\\&&+\sin^2\theta\left[r^2+a^2+a^2\sin^2\theta\left(\frac{2Mr+ \gamma r^{1-3\omega_q}}{\Sigma}\right)\right]d\phi^2\ ,
\end{eqnarray}
where
\begin{eqnarray}\label{04}
\Delta(r)=r^2-2Mr+a^2- \gamma r^{1-3\omega_q},
\end{eqnarray}
and 
\begin{eqnarray}\label{05}
\Sigma=r^2+a^2\cos^2\theta\ .
\end{eqnarray}
We get the horizons of the black hole by simply solving $\Delta{(r)}=0$. For $\omega_q=-2/3$, the equation (\ref{04}) reduces to the cubic polynomial equation, and we find the radius of the horizon by solving
\begin{equation}
    r^2 - 2 M r + a^2 - \gamma r^3 = 0,
\end{equation}
which also can be written as
\begin{equation}
    \gamma(r-r_{in})(r-r_{out})(r-r_{q})=0,
\end{equation}
where $r_{in}$ and $r_{out}$ are the inner and outer horizon, and $r_q$ corresponds to the quintessential  horizon respectively. The rotating black hole in the quintessential dark energy corresponds to three horizons which vary with the spin parameter $a$ and the normalization parameter $\gamma$ as shown in Table~(\ref{table2}). In  Fig.~\ref{fig2}, we show the variations of horizons with the corresponding values of $a$ and $\gamma$.
For extremely rotating case $a=1.15$, the black hole corresponds only quintessential horizon (cf.~Fig.~{\ref{fig2}} and Table~(\ref{table2})).
 {\section{Effective Potential and Impact Parameters}\label{sectwo}}
\begin{figure*}
\begin{center}
    \begin{tabular}{c c}
	\includegraphics[scale=0.45]{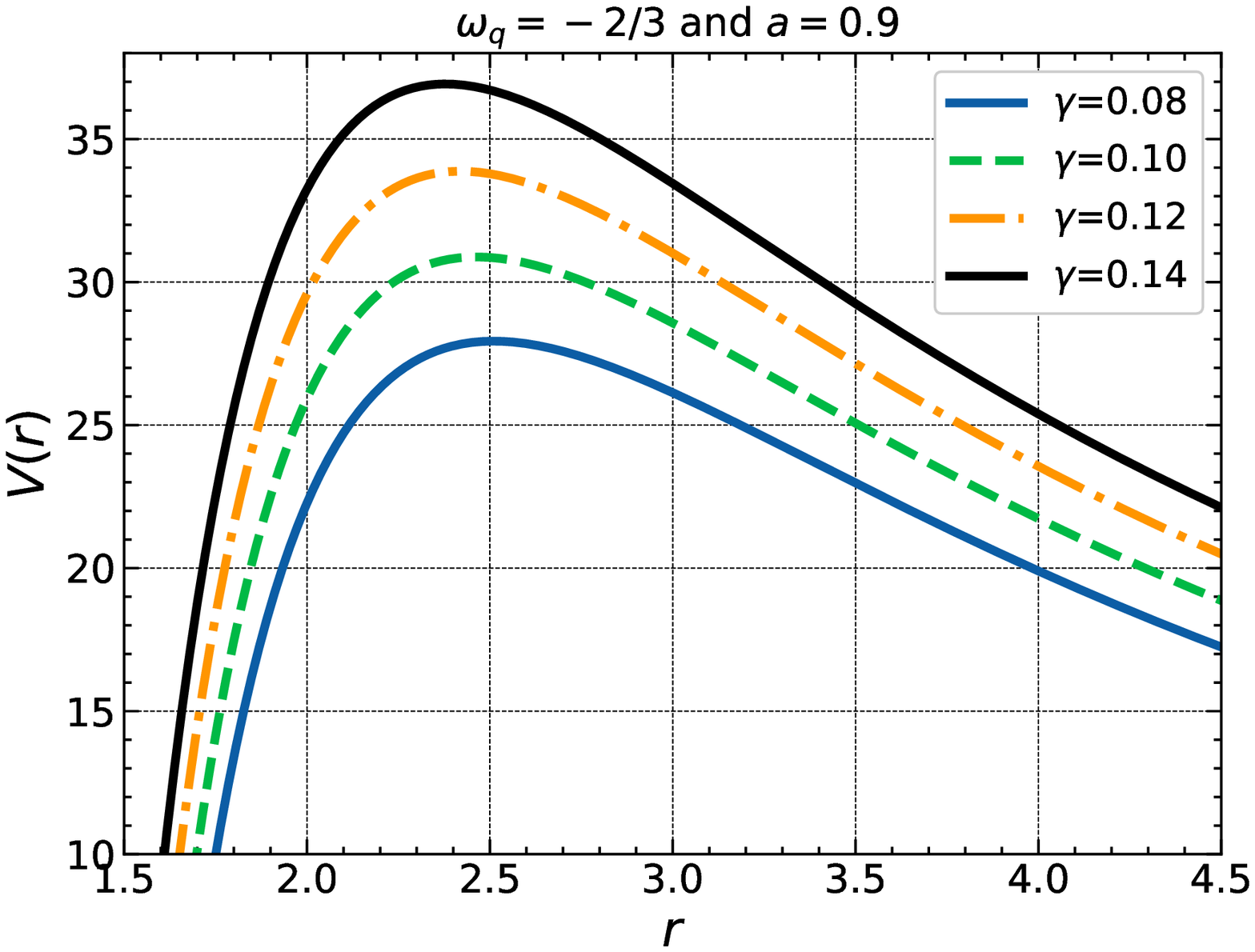} 
	\includegraphics[scale=0.45]{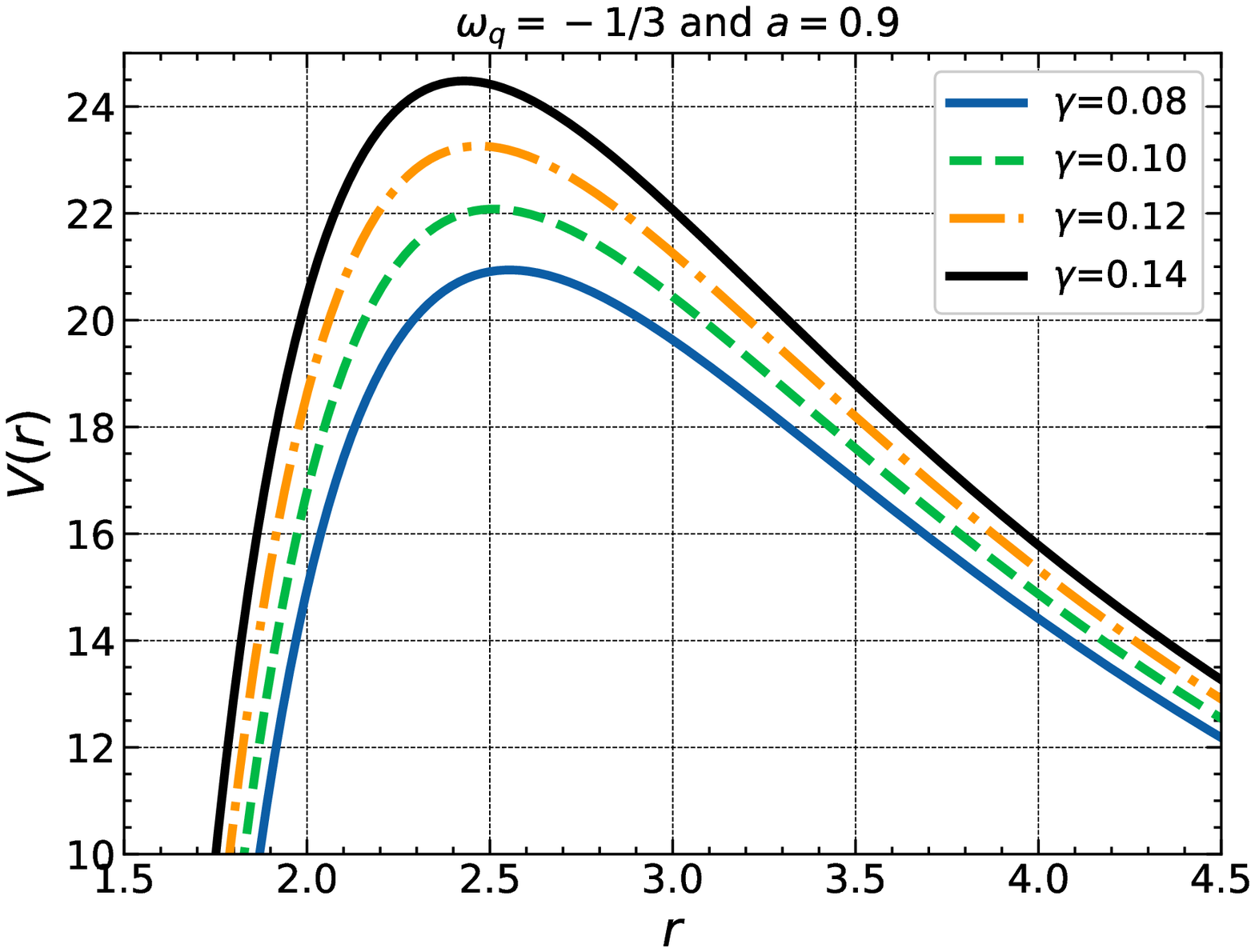}\\
	\includegraphics[scale=0.45]{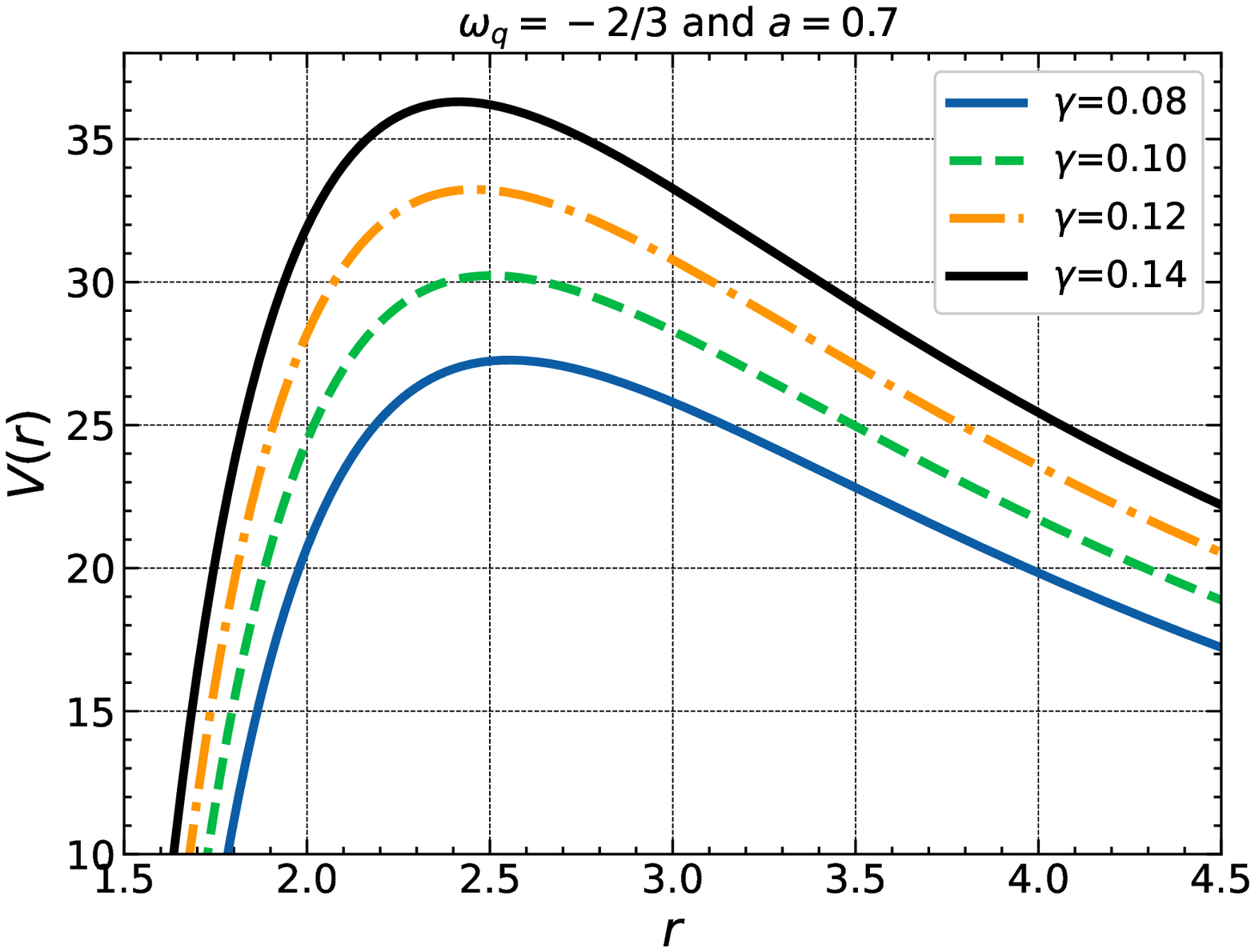}
	\includegraphics[scale=0.45]{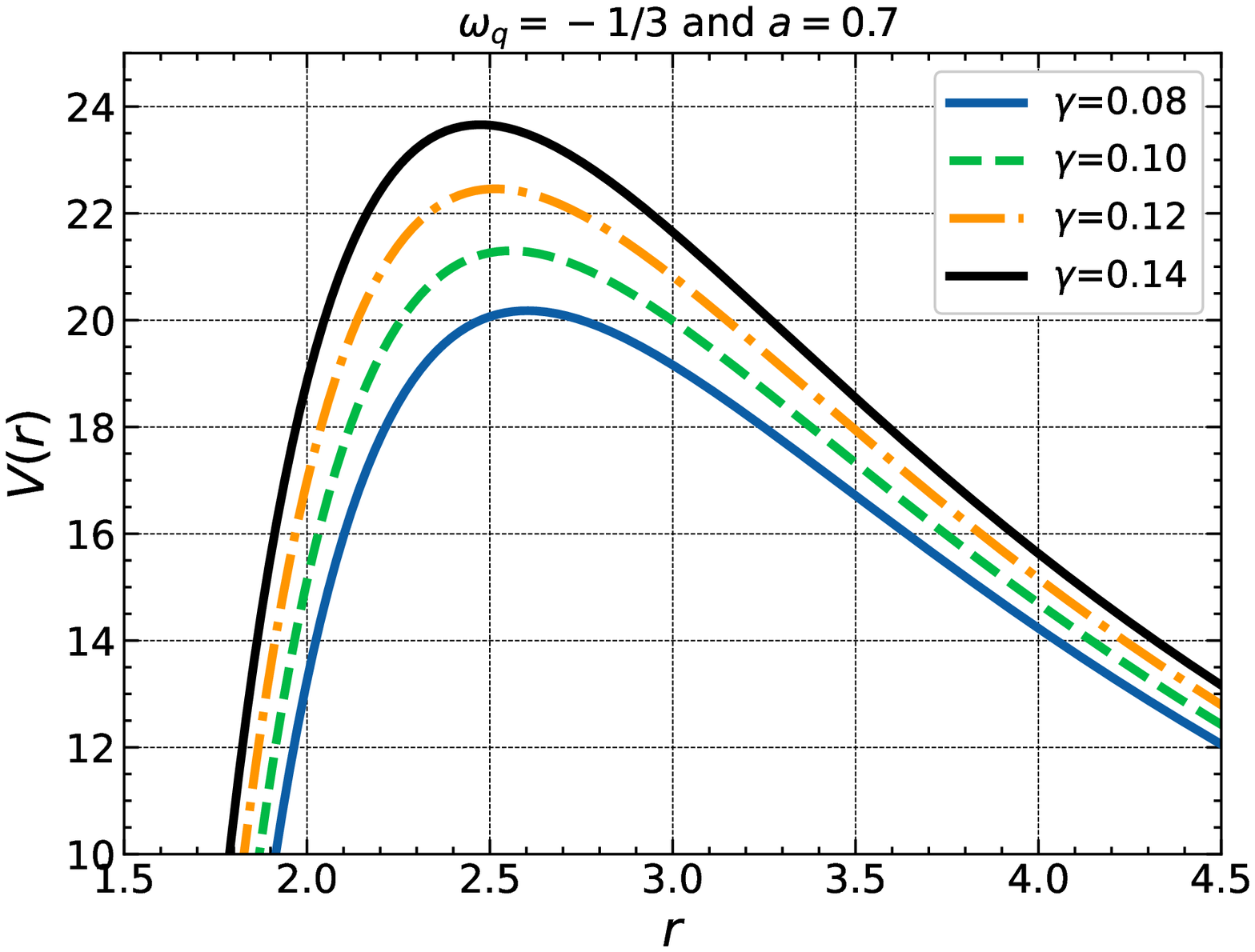}
	 \end{tabular}
    \caption{\label{fig3} Plot showing the variation of effective potential with $r$ for different values of $a$, $\omega_q$ and $\gamma$.}
    \end{center}
\end{figure*}

\begin{figure}
\begin{center}
     \begin{tabular}{c c}
	\includegraphics[scale=0.45]{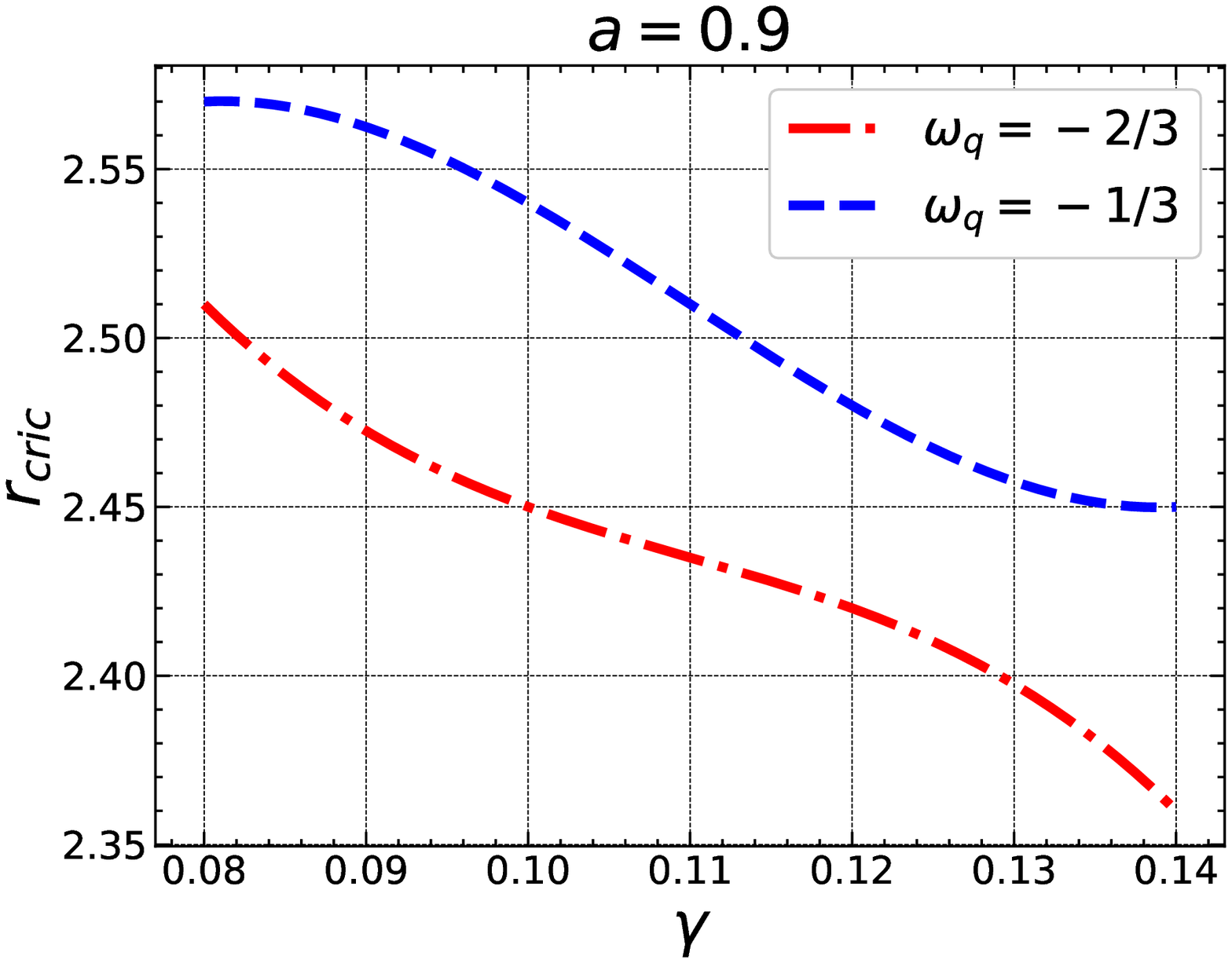} 
	\includegraphics[scale=0.45]{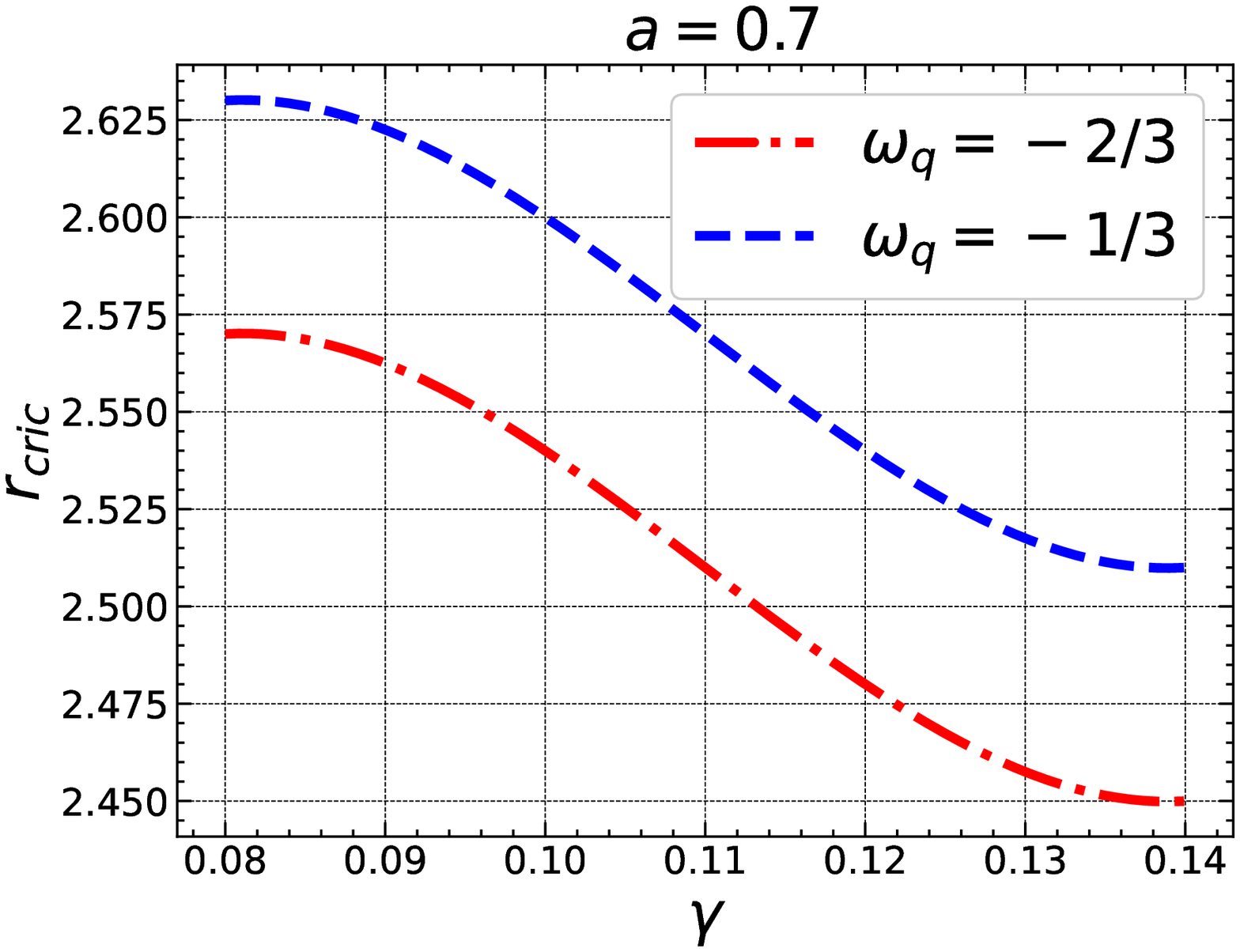} 
	 \end{tabular}
    \caption{\label{fig4} Plot showing the variation of critical radius $r_{cric}$ with $\gamma$ for different values of quintessential parameter $\omega_q$ and spin parameter $a$.}
    \end{center}
\end{figure}

\begin{table}[]
\begin{tabular}{|cccc|cccc|}
\hline
\multicolumn{4}{|c|}{$ a=0.9$} & \multicolumn{4}{c|}{$ a=0.7$} \\ \hline \hline
\multicolumn{2}{|c|}{$\omega_q=-2/3$} & \multicolumn{2}{c|}{$\omega_q=-1/3$} & \multicolumn{2}{c|}{$\omega_q=-2/3$} & \multicolumn{2}{c|}{$\omega_q=-1/3$} \\ \hline
\multicolumn{1}{|c|}{$\gamma$} & \multicolumn{1}{c|}{$r_{cric}$} & \multicolumn{1}{c|}{$\gamma$} & $r_{cric}$ & \multicolumn{1}{c|}{$\gamma$} & \multicolumn{1}{c|}{$r_{cric}$} & \multicolumn{1}{c|}{$\gamma$} & $r_{cric}$ \\ \hline
\multicolumn{1}{|c|}{0.08} & \multicolumn{1}{c|}{2.51} & \multicolumn{1}{c|}{0.08} & 2.57 & \multicolumn{1}{c|}{0.08} & \multicolumn{1}{c|}{2.57} & \multicolumn{1}{c|}{0.08} & 2.63 \\ \hline
\multicolumn{1}{|c|}{0.10} & \multicolumn{1}{c|}{2.45} & \multicolumn{1}{c|}{0.10} & 2.54 & \multicolumn{1}{c|}{0.10} & \multicolumn{1}{c|}{2.54} & \multicolumn{1}{c|}{0.10} & 2.60 \\ \hline
\multicolumn{1}{|c|}{0.12} & \multicolumn{1}{c|}{2.42} & \multicolumn{1}{c|}{0.12} & 2.48 & \multicolumn{1}{c|}{0.12} & \multicolumn{1}{c|}{2.48} & \multicolumn{1}{c|}{0.12} & 2.54 \\ \hline
\multicolumn{1}{|c|}{0.14} & \multicolumn{1}{c|}{2.36} & \multicolumn{1}{c|}{0.14} & 2.45 & \multicolumn{1}{c|}{0.14} & \multicolumn{1}{c|}{2.45} & \multicolumn{1}{c|}{0.14} & 2.51 \\ \hline
\end{tabular}
\caption{\label{table3}Table showing the variation of $r_{cric}$ with $a$, $\omega_{q}$ and $\gamma$}
\end{table}
The motion of particles and electromagnetic radiations around the rotating spacetime is regular so the equations of motion of a particle around a black hole in  quintessential dark energy are completely integrable. In this section, we derive complete integral equations of motion and study the effective potential of the black hole and calculate the impact parameters. We use  the Hamilton-Jacobi formalism to find the complete geodesic equations which were originally derived by Carter \cite{wheeler,novikov,Chandrasekhar:1992,Carter:1968rr}. The Hamilton-Jacobi equation reads
\begin{eqnarray}
\label{eqn8}
\frac{\partial S}{\partial \upsilon} = -\frac{1}{2}g^{\mu\nu} p_\mu p_\nu ,
\end{eqnarray}
where $\upsilon$ is the affine parameter and $p_{\mu}$ is the conjugate momenta corresponding to the first derivative of Jacobian action $S$ with respect to the generalized coordinate. We choose a separable solution for the Jacobean action in the given form
\begin{eqnarray}
\label{eqn9}
S=\frac12 {m_{0}}^2 \upsilon -{\cal E} t +{\cal L} \phi + S_r(r)+S_\theta(\theta) ,
\end{eqnarray}
where $m_0$ is the rest mass of the test particle which is zero for the photon. The black hole metric is cyclic for time $t$ and azimuthal $\phi$ coordinates. There exist two conserved quantities corresponding to these cyclic coordinates $\cal{E}$ and $\cal{L}$ having units of energy and angular momentum respectively. Using our separable solution defined in Eq.~(\ref{eqn9}), we obtain our complete  equations of motion for a massive test particle in the first-order differential form as
\begin{eqnarray}
\Sigma \frac{dt}{d\upsilon}&=&\frac{r^2+a^2}{\Delta(r)}\left[{\cal E}(r^2+a^2)-a{\cal L}\right]  -a(a{\cal E}\sin^2\theta-{\mathcal {L}})\ ,\label{t}\\
\Sigma \frac{dr}{d\upsilon}&=&\sqrt{\mathcal{R}(r)}\ ,\label{r}\\
\Sigma \frac{d\theta}{d\upsilon}&=&\sqrt{\Theta(\theta)}\ ,\label{th}\\
\Sigma \frac{d\phi}{d\upsilon}&=&\frac{a}{\Delta(r)}\left[{\cal E}(r^2+a^2)-a{\cal L}\right]-\left(a{\cal E}-\frac{{\cal L}}{\sin^2\theta}\right)\ ,\label{phi}
\end{eqnarray}
where 
\begin{eqnarray}
&&\mathcal{R}(r)=\left[(r^2+a^2){\cal E}-a{\cal L}\right]^2-\Delta(r)\left[{m_0}^2r^2+(a{\cal E}-{\cal L})^2+{\cal K}\right],\\
&&\Theta(\theta)={\cal K}-\left[\frac{{\cal L}^2}{\sin^2\theta}-a^2 {\cal E}^2\right]\cos^2\theta\ .
\end{eqnarray}
\begin{figure*}
\begin{center}
    \begin{tabular}{c c }
	\includegraphics[scale=0.6]{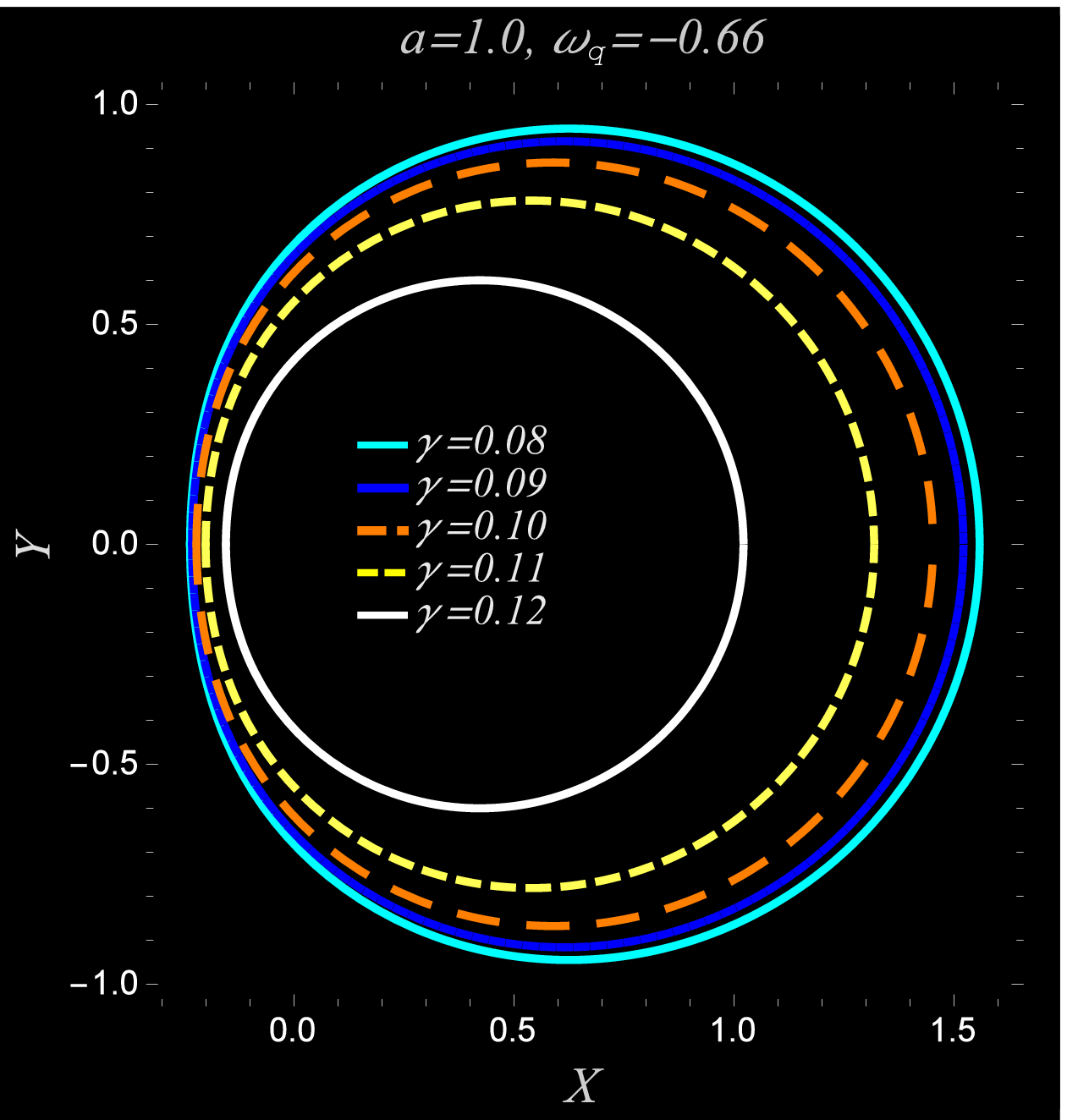} 
	\includegraphics[scale=0.61]{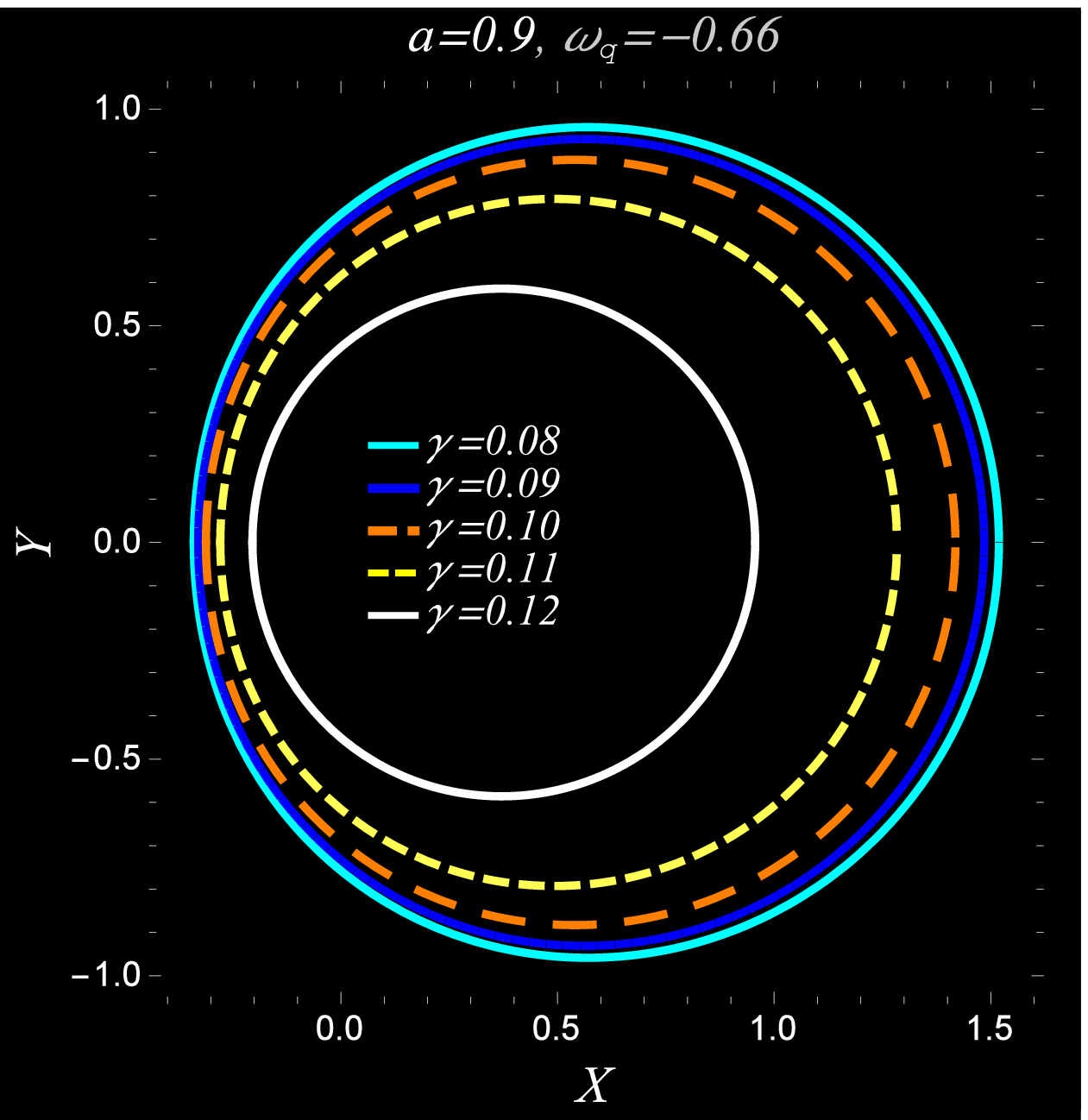}\\
	\includegraphics[scale=0.6]{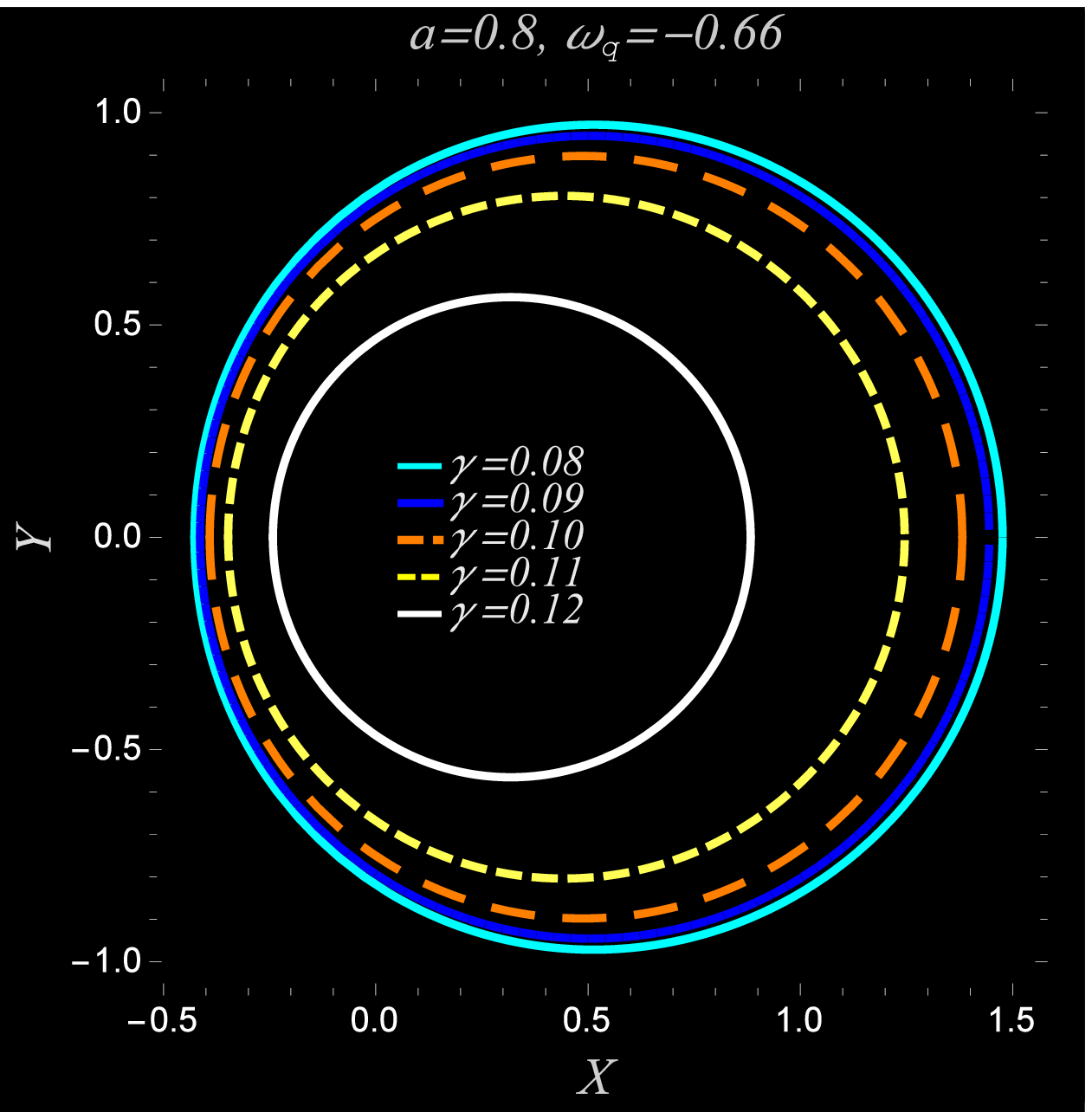}
	\includegraphics[scale=0.6]{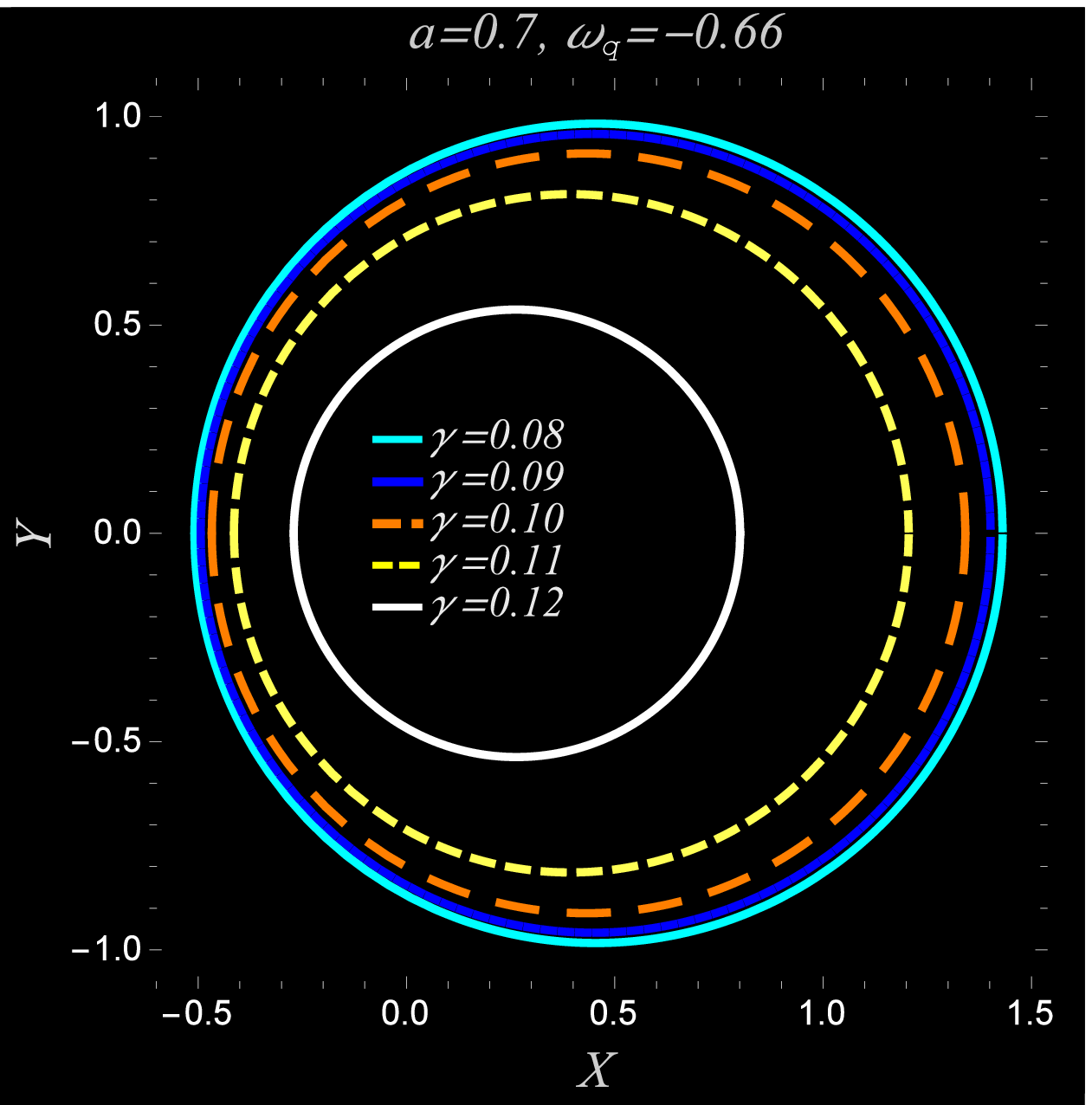} 
	 \end{tabular}
    \caption{\label{fig5}  Plot showing the shadows of the rotating black holes in quintessential dark energy for the different values of the parameter $\gamma$, $a$.}
    \end{center}
\end{figure*}
\begin{figure*}
\begin{center}
    \begin{tabular}{c c }
	\includegraphics[scale=0.6]{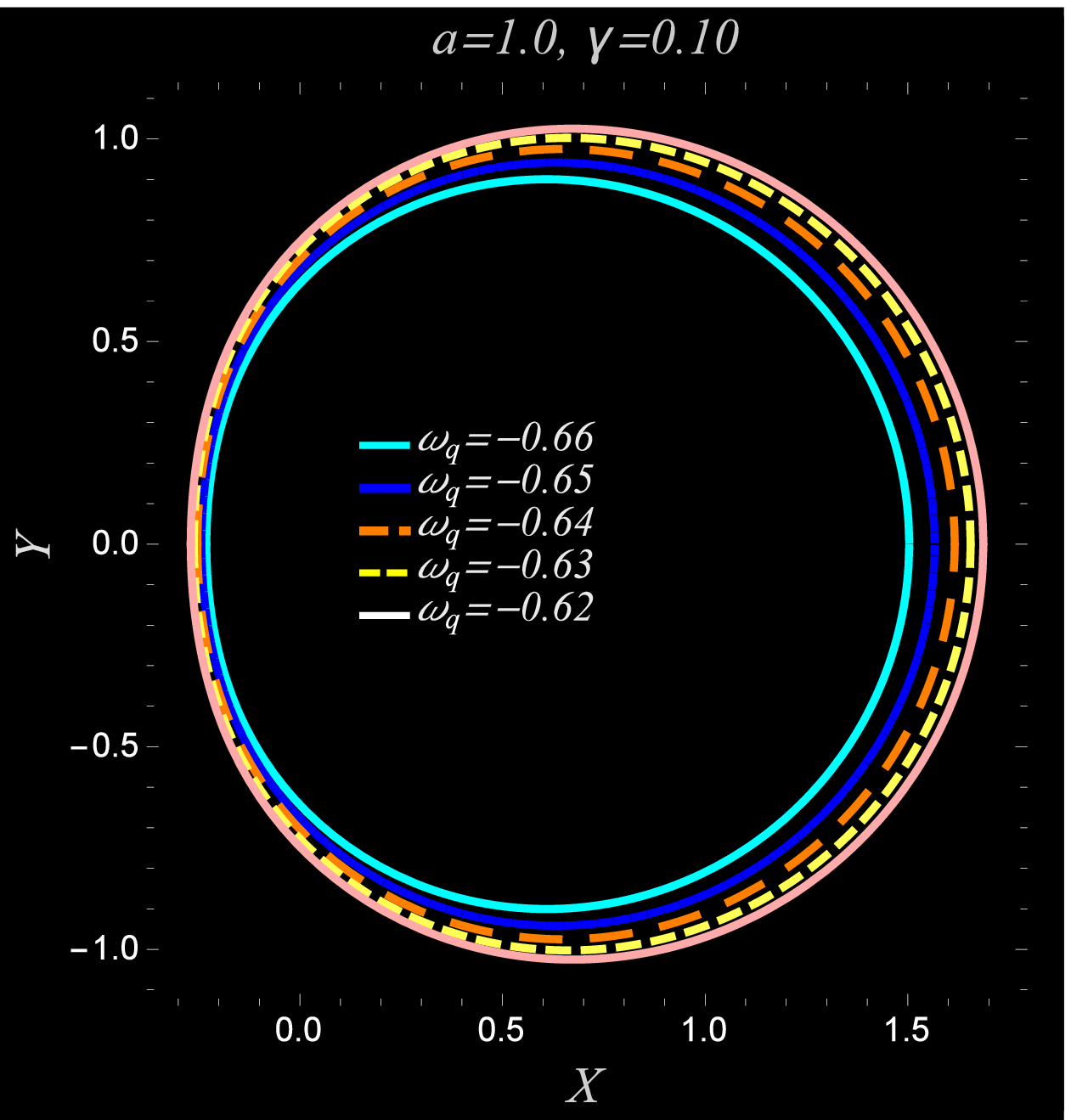} 
	\includegraphics[scale=0.61]{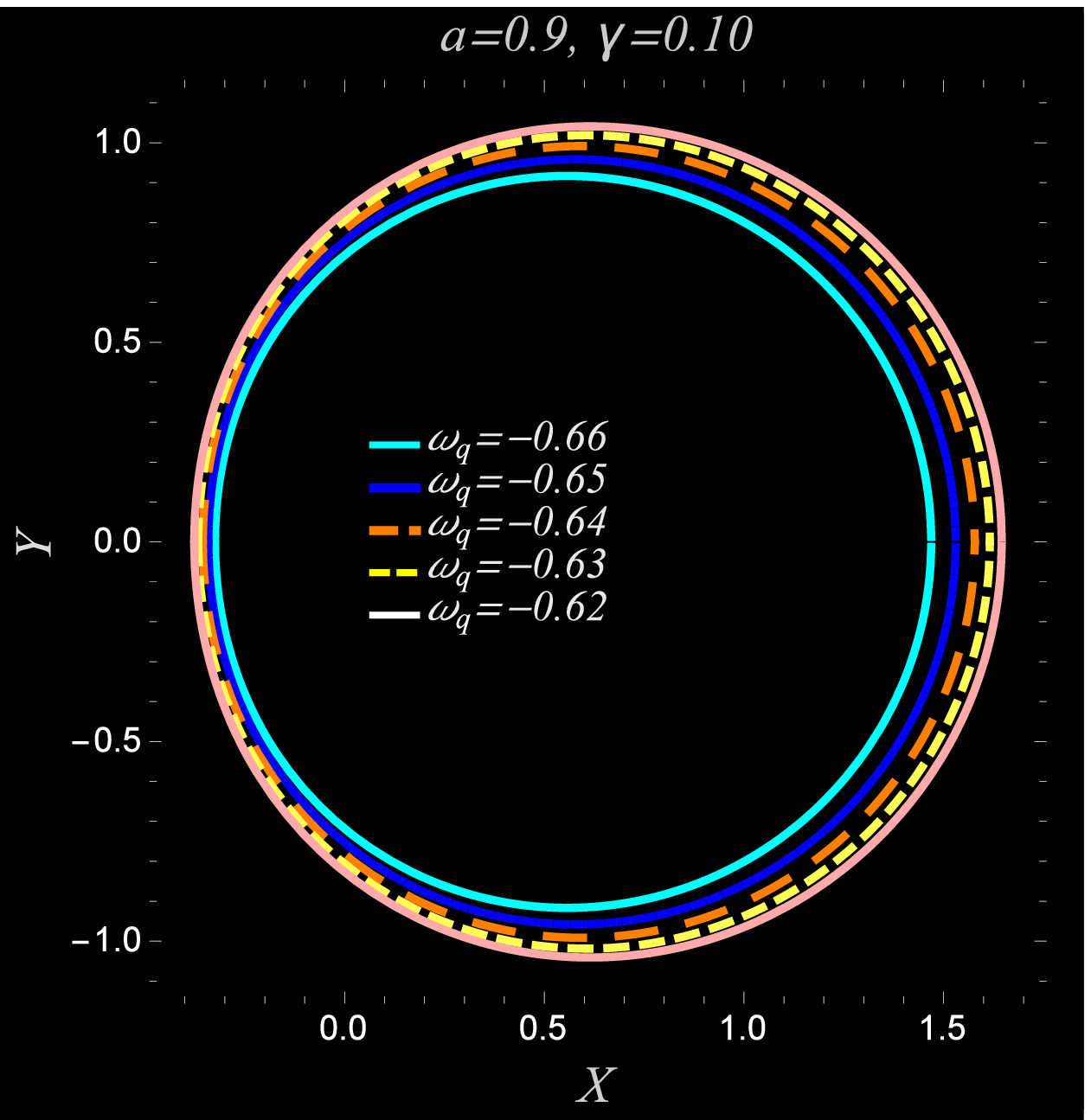}\\
	\includegraphics[scale=0.6]{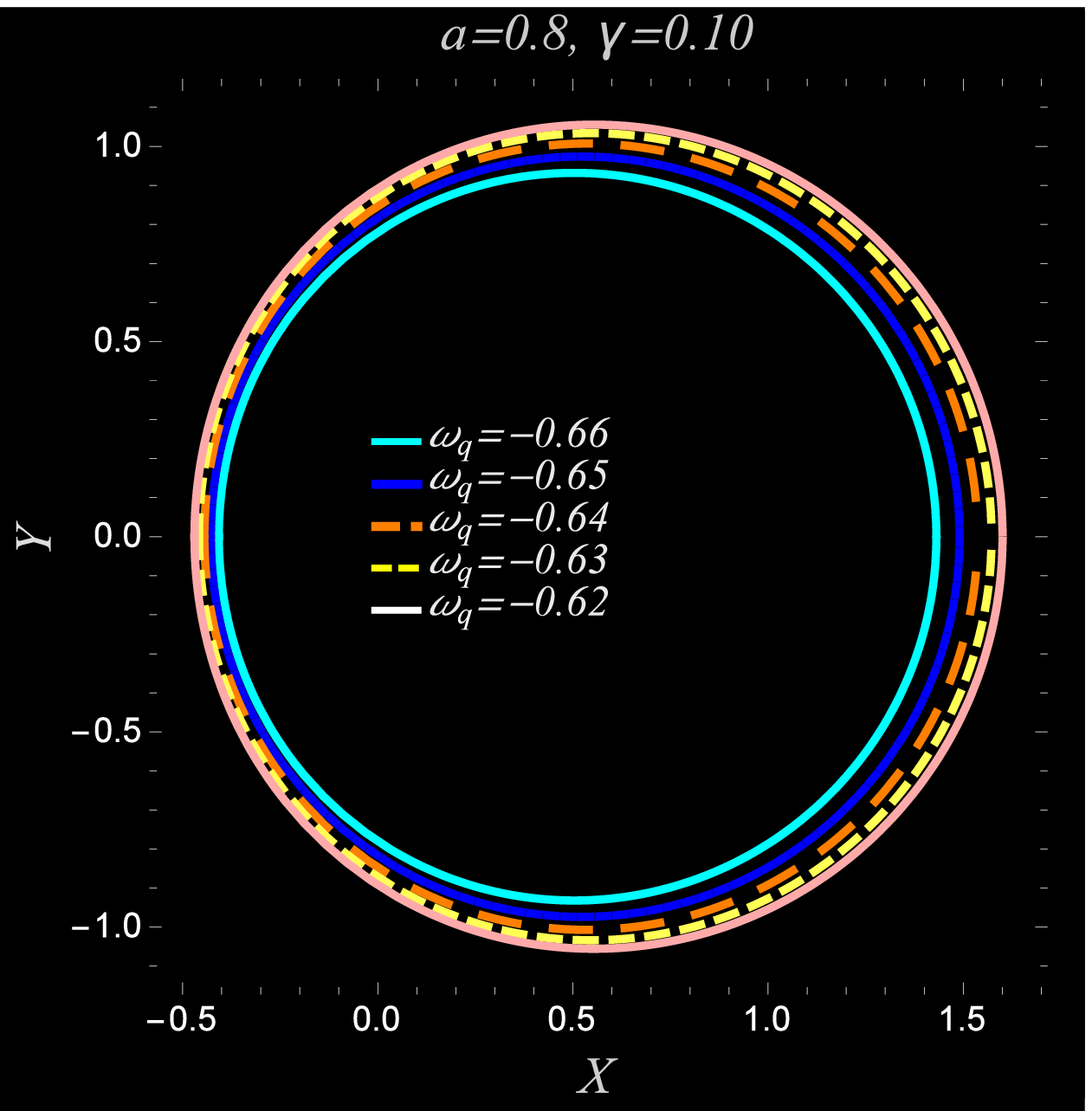} 
	\includegraphics[scale=0.6]{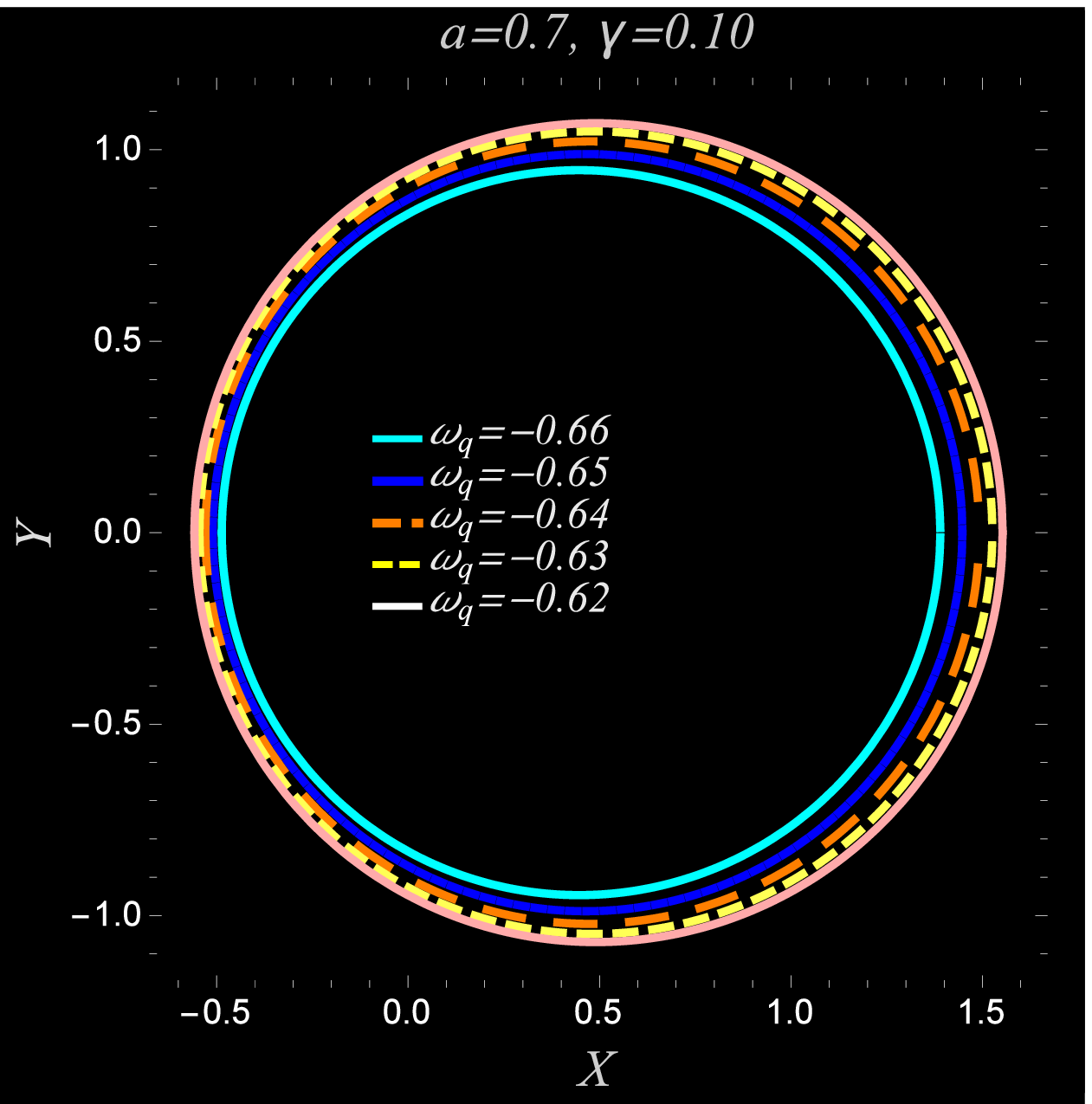}
	 \end{tabular}
    \caption{\label{fig6}  Plot showing the shadows of the rotating black holes in quintessential dark energy for the different values of the parameter $\omega_q$, $a$  .}
    \end{center}
\end{figure*}
Here, $\mathcal{K}$ is the Carter separable constant. The dynamics of a test particle around the black hole can be obtained from the geodesic Eqs.~(\ref{t})-(\ref{phi}). These geodesic equations reduce for the Kerr black holes in the absence of the quintessential field $\gamma=0$  \cite{novikov}. The study of the motion of the test particle around the rotating black hole requires two impact parameters. Here, we are introducing two impact parameters $\eta$ and $\xi$ in terms of conserved quantities $\mathcal{E}$, $\mathcal{L}$ and $\mathcal{K}$ as
\begin{equation}
\xi=\mathcal{L}/\mathcal{E}, \quad\quad \eta=\mathcal{K}/\mathcal{E}^2.
\end{equation}
 The rest mass of the photon is zero, so for the case of null geodesics, we consider $m_{0}=0$. The radial equation of motion   in terms of  impact parameters reads
\begin{equation}
\mathcal{R}(r)=\frac{1}{{\cal{E}}^2}\left[[(r^2+a^2) -a{\xi}]^2-\Delta(r)[(a -{\xi})^2+{\eta}]\right].
\label{17}
\end{equation}
\begin{figure*}
\begin{center}
    \begin{tabular}{c c }
	\includegraphics[scale=0.65]{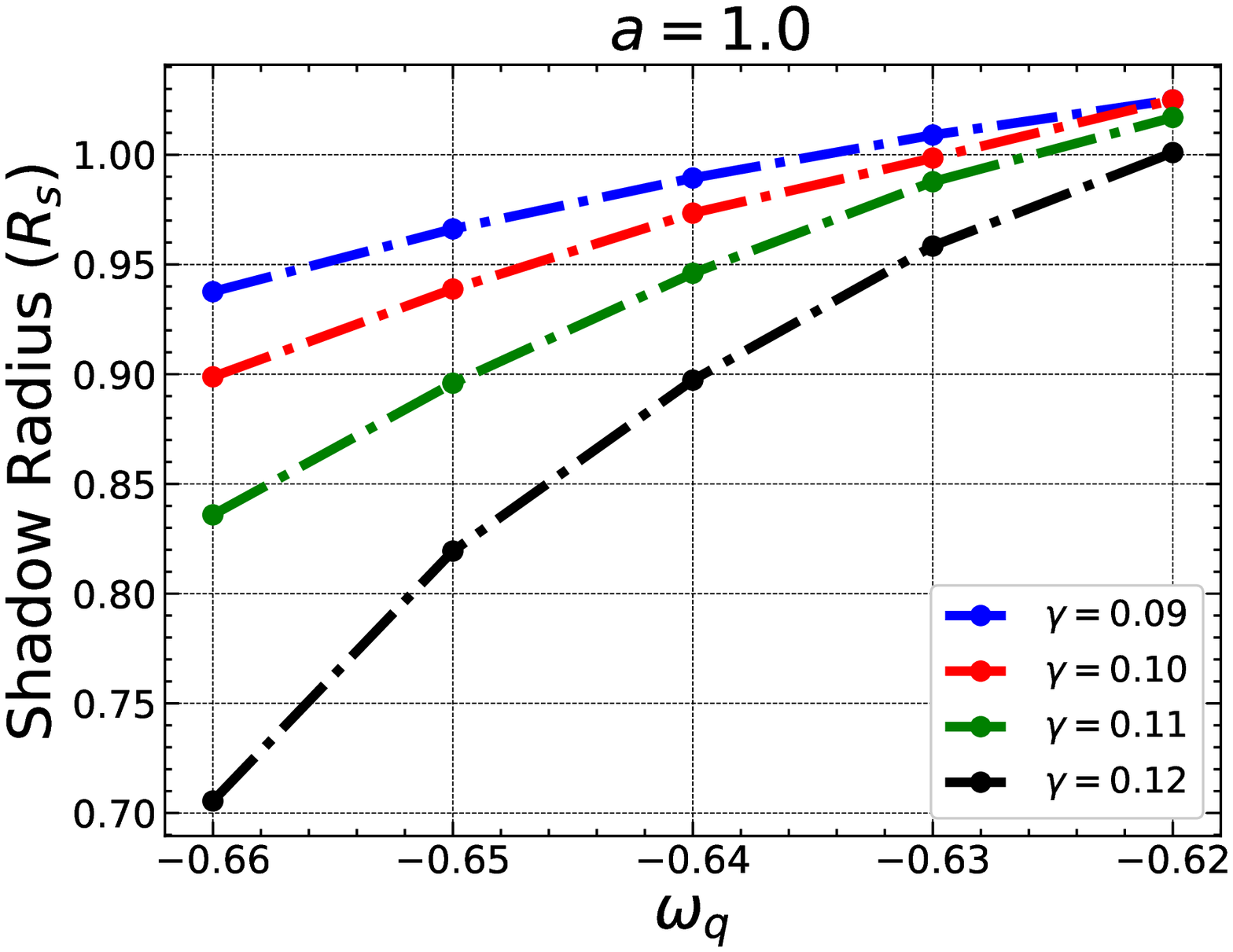} 
	 \end{tabular}
    \caption{\label{fig7} Plot showing the variation of shadow radius $R_s$ with the quintessential dark energy parameter $\omega_q$ for different $\gamma$.}
    \end{center}
\end{figure*}
The formation of circular orbits around the black hole is an important feature for the study of gravitational lensing and shadow properties. The idea of the formation of  circular orbits can be obtained from the effective potential of the black hole. The circular photon orbits  may be categorized into two types: unstable or stable circular photon orbits. When a ray of photon departs from the unstable circular orbit then it may fall into the black hole or reach  the infinite observer. No bound orbits exit around the unstable circular orbits. Instead of that, many possible bound orbits  may exist around the stable circular orbits. We use the radial equation of motion (\ref{17}) to find the expression of the effective potential.  The radial equation of motion can be rewritten in terms of effective potential as $V_{eff}$ as
\begin{equation}\label{veff}
\left(\frac{d{r}}{d\upsilon}\right)^2+V_{eff}(r)=0,
\end{equation}
from the above equation (\ref{veff}), one can find the expression of the effective potential which takes the following form
\begin{equation}
V_{eff}=\frac{1}{{{\Sigma}^2}}[ ((r^2+a^2) -a{\xi})^2-\Delta(r) ( (a -{\xi})^2+{\eta}) ].\label{vef}
\end{equation}
In Fig.~(\ref{fig3}), we plot the behaviour of effective potential with radial coordinate $r$ for different values of $a$, $ \omega_q $ and $\gamma$. One can observe from  Fig.~(\ref{fig3}) that for a fixed value of $a$ and $\omega_q$ the maximum value or peak value increases with the increasing value of $\gamma$. Another interesting result one can observe is that the critical radius corresponding to the peak value of effective potential shifts to the left which means the radii of the unstable circular orbits decreases with the increasing values of $\gamma$ (cf. Fig.~(\ref{fig4}) and Table \ref{table3}). The critical radius $r_{cri}$ of unstable photon orbits corresponds to the local maximum of the effective potential while the critical radius  for the most stable photon orbits corresponds to the local minimum.  Here we are interested to find the most unstable photon orbits and for that, we must have to maximize the effective potential which satisfies the following condition
\begin{equation}
V_{eff}=\frac{\partial V_{eff}}{\partial r}=0 \;\; \;\; \mbox{or}\;\;  \; \mathcal{R}=\frac{\partial \mathcal{R}}{\partial r}=0,\label{20} 
\end{equation}
by applying the above conditions (\ref{20}),  we obtain the individual expressions for impact parameters
 \begin{eqnarray}
 \eta&=&\frac{1}{a^2 \left(-3 \gamma \omega _q+\gamma+2 (M-r) r^{3 \omega _q}\right){}^2}\Big[r^3(-4 \left(r (r-3 M)^2-4 a^2 M\right) r^{6 \omega _q})\nonumber\\
 &+&4 \gamma r^{3 \omega _q} \left(3 \omega _q \left(2 a^2+r (r-3 M)\right)+2 a^2+3 r (r-3 M)-9 {\gamma}^2 r \left(\omega _q+1\right){}^2\right)\Big]\nonumber,\\
 \end{eqnarray}
\begin{eqnarray}
\xi&=&\frac{1}{a \left(-3 \gamma \omega _q+\gamma+2 (M-r) r^{3 \omega _q}\right)}\Big[2 \left(a^2 (M+r)+r^2 (r-3 M)\right) r^{3 \omega _q}\nonumber\\
&+& \gamma \left(-3 \left(a^2+r^2\right) \omega _q+a^2-3 r^2\right)\Big].
\end{eqnarray}
\begin{figure*}
\begin{center}
    \begin{tabular}{c c }
	\includegraphics[scale=0.65]{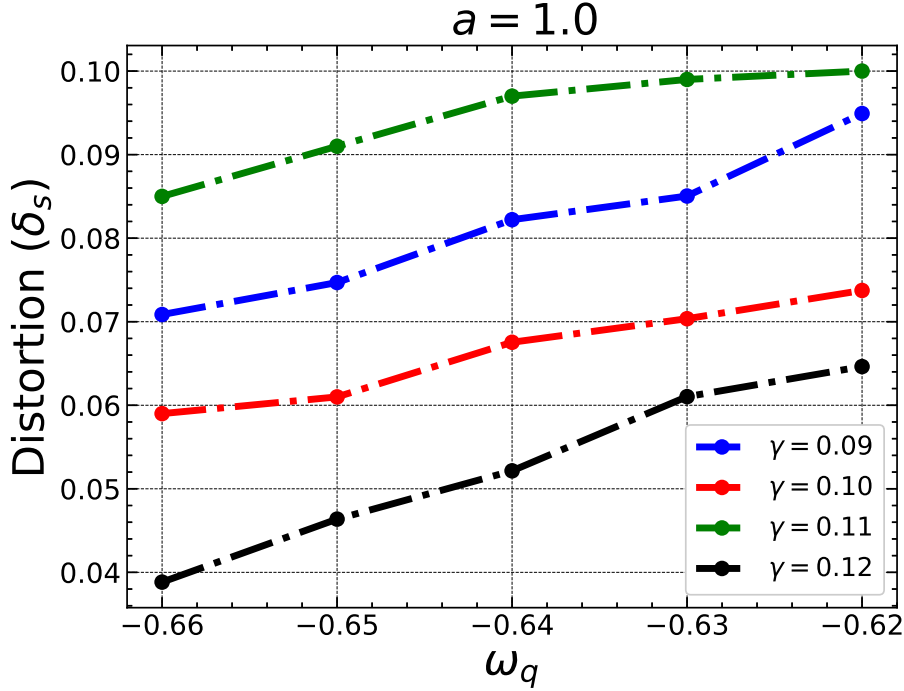}\\
	 \end{tabular}
    \caption{\label{fig8} Plot showing the variation of the distortion parameter with the quintessential dark energy parameter $\omega_q$ for different parameter $\gamma$ .}
    \end{center}
\end{figure*}
These expressions of $\eta$ and $\xi$ define the motion of photons around the black holes in quintessential dark energy. These equations of impact parameters exactly reduce for the Kerr black holes in the absence of the quintessence  dark energy field  \cite{novikov}.
\section{Black hole shadow}\label{secthree}
A black hole form shadow due to the infalling light rays inside its event horizon. The light rays at the unstable equilibrium form orbit around it. We consider orthogonal tetrad to define the celestial coordinate along the boundary of the black hole shadow. The orthonormal tetrads are 
\begin{eqnarray}
e_0=\frac{(r^2+a^2)\partial_t+a\partial_{\phi}}{\sqrt{\Sigma\Delta}},\;\;\;\;e_1=\frac{1}{\sqrt{\Sigma}}\partial_{\theta},\nonumber\\
e_2=-\frac{\partial_{\phi}+a\sin^2{\theta}\partial_t}{\sqrt{\Sigma \Delta}},\;\;\;\; e_3=-\sqrt{\frac{\Delta}{\Sigma}}\partial_r.
\end{eqnarray} 
The tangent vector   over a light ray  can be defined as
\begin{equation}
\dot{\lambda}(\upsilon)=\dot{t}\partial_t+\dot{r}\partial_r+\dot{\theta}\partial_{\theta}+\dot{\phi}\partial_{\phi},\label{24}
\end{equation}
and the tangent vector for the observers position is
\begin{equation}
\dot{\lambda}(\upsilon)=\alpha(-e_0+\sin\Phi\cos\Psi e_1 +\sin\Phi\sin\Psi e_2 +\cos\Phi e_3 ).\label{25}
\end{equation} 
The expression for the scalar factor $\alpha$ can be  determined by comparing Eq.~($\ref{24}$) and (\ref{25}), and using Eq.~(\ref{t} - \ref{phi}), which takes the following form 
\begin{equation}
\alpha=\frac{a\mathcal{L}-(r^2+a^2)\mathcal{E}}{\sqrt{\Sigma\Delta}}.\label{alpha1}
\end{equation}
Next, the celestial coordinates $\phi$ and $\psi$ can be defined  as
\begin{eqnarray}
\sin\Psi(r_O,r_p) &=&\left.\left( \frac{\xi-a}{\sqrt{(a-\xi)^2+\eta}}\right)\right|_{r=r_O},\\ \label{Psieq}
\sin\Phi(r_O,r_p) &=&\left.\left(\frac{\sqrt{\Delta[(a-\xi)^2+\eta]}}{(r^2+a^2-a\xi)}\right)\right|_{r=r_O},\label{Phieq}
\end{eqnarray}
where the impact parameters $\xi$ and $\eta$ are functions of the radius of the unstable photon orbits around the black hole. To visualize a shadow, we introduce the Cartesian coordinates $X$ and $Y$ \cite{Grenzebach:2014fha}
\begin{eqnarray}
X(r_O,r_p)=-2\tan\left(\frac{\Phi(r_O,r_p)}{2}\right)\sin(\Psi(r_O,r_p)), \label{X11}\nonumber\\
Y(r_O,r_p)=-2\tan\left(\frac{\Phi(r_O,r_p)}{2}\right)\cos(\Psi(r_O,r_p)),\label{Y1}
\end{eqnarray}  
which satisfies the following relation
\begin{equation}
X^2+Y^2=4\tan^{2}\Big(\frac{\Phi(r_{o},r_{p})}{2}\Big).
\end{equation}
The contour plot of $X$ vs $Y$ defines the photon region or shadow boundary of rotating black hole in quintessential dark energy as seen by static observer at the distance $r_{O}=5M$ in the domain of outer communication. We plot various figures for the black hole shadow in quintessence for different values of spin parameter $a$, state parameter $\omega_q$ and normalization factor $\gamma$ (cf. Fig.~\ref{fig5} and \ref{fig6}). The effective size of the black hole shadow decreases with $\omega_q$ (Fig.~\ref{fig5}) while the shadow size increases for the increasing values of $\gamma$ (Fig.~\ref{fig6}). The next important result one can observe from the shadow plots, the shape of the black hole shadow distorted with the quintessence parameter $\omega_q$ while for the increasing values of normalization factor $\gamma$ the black hole shadow appears more circular even for the very high values of spin parameter which we have clearly shown in the first plot of the figure~\ref{fig5}.
\begin{table}[]
\begin{tabular}{|c|cccc|cccc|}
\hline
\multirow{2}{*}{$\omega_q$} &
  \multicolumn{4}{c|}{$\gamma=0.09$} &
  \multicolumn{4}{c|}{$\gamma=0.10$} \\ \cline{2-9} 
 &
  \multicolumn{1}{c|}{$R_s$} &
  \multicolumn{1}{c|}{$\delta_s$} &
  \multicolumn{1}{c|}{$A$} &
  $\theta_d$ &
  \multicolumn{1}{c|}{$R_s$} &
  \multicolumn{1}{c|}{$\delta_s$} &
  \multicolumn{1}{c|}{$A$} &
  $\theta_d$ \\ \hline
-0.66 &
  \multicolumn{1}{c|}{0.9376} &
  \multicolumn{1}{c|}{0.07086} &
  \multicolumn{1}{c|}{2.7617000} &
  21.489792 &
  \multicolumn{1}{c|}{0.8988} &
  \multicolumn{1}{c|}{0.05900} &
  \multicolumn{1}{c|}{2.53790873} &
  20.600496 \\ \hline
-0.65 &
  \multicolumn{1}{c|}{0.9662} &
  \multicolumn{1}{c|}{0.07469} &
  \multicolumn{1}{c|}{2.93281007} &
  22.145304 &
  \multicolumn{1}{c|}{0.9388} &
  \multicolumn{1}{c|}{0.06100} &
  \multicolumn{1}{c|}{2.76882836} &
  21.517296 \\ \hline
-0.64 &
  \multicolumn{1}{c|}{0.9894} &
  \multicolumn{1}{c|}{0.08221} &
  \multicolumn{1}{c|}{3.07534388} &
  22.677048 &
  \multicolumn{1}{c|}{0.9734} &
  \multicolumn{1}{c|}{0.06754} &
  \multicolumn{1}{c|}{2.97668279} &
  22.310328 \\ \hline
-0.63 &
  \multicolumn{1}{c|}{1.0090} &
  \multicolumn{1}{c|}{0.08504} &
  \multicolumn{1}{c|}{3.19839579} &
  23.12628 &
  \multicolumn{1}{c|}{0.9985} &
  \multicolumn{1}{c|}{0.07036} &
  \multicolumn{1}{c|}{3.13217494} &
  22.88562 \\ \hline
-0.62 &
  \multicolumn{1}{c|}{1.0250} &
  \multicolumn{1}{c|}{0.09492} &
  \multicolumn{1}{c|}{3.30063578} &
  23.49301 &
  \multicolumn{1}{c|}{1.0250} &
  \multicolumn{1}{c|}{0.07374} &
  \multicolumn{1}{c|}{3.30063578} &
  23.49300 \\ \hline
\multirow{2}{*}{$\omega_q$} &
  \multicolumn{4}{c|}{$\gamma=0.11$} &
  \multicolumn{4}{c|}{$\gamma=0.12$} \\ \cline{2-9} 
 &
  \multicolumn{1}{c|}{$R_s$} &
  \multicolumn{1}{c|}{$\delta_s$} &
  \multicolumn{1}{c|}{$A$} &
  $\theta_d$ &
  \multicolumn{1}{c|}{$R_s$} &
  \multicolumn{1}{c|}{$\delta_s$} &
  \multicolumn{1}{c|}{$A$} &
  $\theta_d$ \\ \hline
-0.66 &
  \multicolumn{1}{c|}{0.8359} &
  \multicolumn{1}{c|}{0.08500} &
  \multicolumn{1}{c|}{2.1951213} &
  19.158828 &
  \multicolumn{1}{c|}{0.7055} &
  \multicolumn{1}{c|}{0.03883} &
  \multicolumn{1}{c|}{1.5636657} &
  16.17006 \\ \hline
-0.65 &
  \multicolumn{1}{c|}{0.8959} &
  \multicolumn{1}{c|}{0.09100} &
  \multicolumn{1}{c|}{2.52155791} &
  20.534028 &
  \multicolumn{1}{c|}{0.8194} &
  \multicolumn{1}{c|}{0.04637} &
  \multicolumn{1}{c|}{2.1093167} &
  18.780648 \\ \hline
-0.64 &
  \multicolumn{1}{c|}{0.9460} &
  \multicolumn{1}{c|}{0.09700} &
  \multicolumn{1}{c|}{2.81146153} &
  21.68232 &
  \multicolumn{1}{c|}{0.8973} &
  \multicolumn{1}{c|}{0.05215} &
  \multicolumn{1}{c|}{2.52944481} &
  20.566116 \\ \hline
-0.63 &
  \multicolumn{1}{c|}{0.9877} &
  \multicolumn{1}{c|}{0.09900} &
  \multicolumn{1}{c|}{3.06478477} &
  22.638084 &
  \multicolumn{1}{c|}{0.9584} &
  \multicolumn{1}{c|}{0.06103} &
  \multicolumn{1}{c|}{2.88564886} &
  21.966528 \\ \hline
-0.62 &
  \multicolumn{1}{c|}{1.0170} &
  \multicolumn{1}{c|}{0.1000} &
  \multicolumn{1}{c|}{3.24931472} &
  23.30964 &
  \multicolumn{1}{c|}{1.0010} &
  \multicolumn{1}{c|}{0.06463} &
  \multicolumn{1}{c|}{3.14787898} &
  22.942920 \\ \hline
\end{tabular}
\caption{\label{table4} Table Showing the black hole shadow observables: shadow radius $R_s$, distortion $\delta_s$, Area $A$ and angular diameter $\theta_d$ for spin parameter $a=1.0$  with different $\omega_q$ and $\gamma$. }
\end{table}
 
 \section{Observables}\label{secfour}
  \begin{figure}
 \includegraphics[scale=0.65]{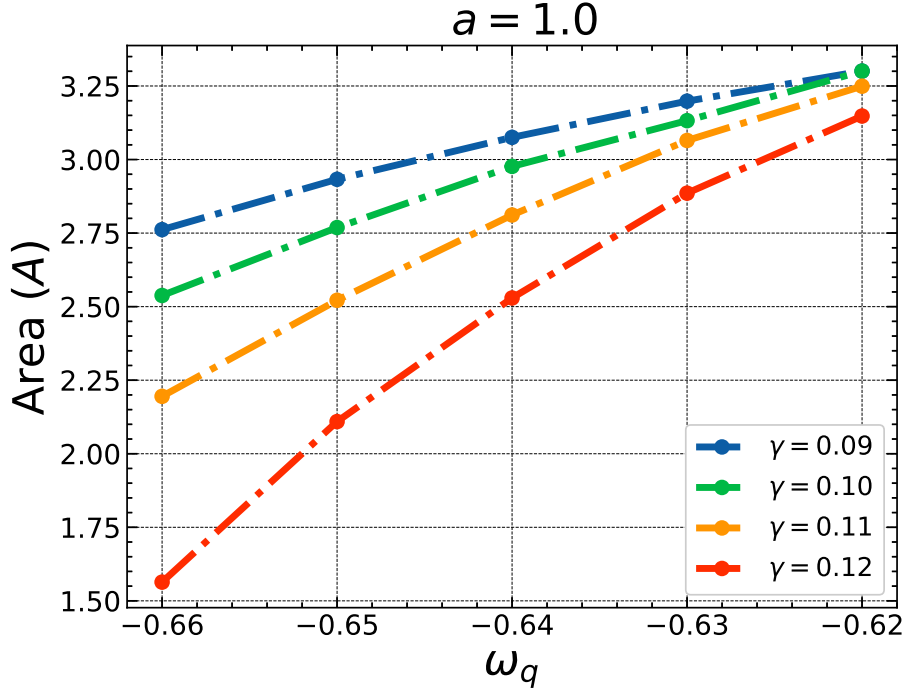}
  \caption{\label{fig9} Plot showing the variation of the black hole shadow area $A$ with the quintessential  dark energy parameter $\omega_q$ }
 \end{figure}
 In this section we estimate observables: shadow radius $R_s$, distortion parameter $\delta_S$ respectively,  which is  originally shared by Hioki and Maeda in \cite{Hioki:2009na}. 
 The observable shadow radius $R_s$ is given by
 \begin{equation}
     R_{s}=\frac{(X_t-X_r)^2 + Y_{t}^2}{2|X_{r}-X_{t}|},
 \end{equation}
 where $(X_t, Y_t)$, $(X_r, Y_r)$ are the top most and right most co-ordinates of the $X$ and $Y$ axis in the Fig.~{\ref{fig5}} and {\ref{fig6}}.  Another observable which is  distortion parameter reads in the form
 \begin{equation}
     \delta=\frac{D}{R_{s}},
 \end{equation}
 where $D=|X_l-X_l^{'}|$ with $X_l$ and $X_l^{'}$ are the left most co-ordinates on the $X$ axis from where the black hole shadow and reference circle passes.

In Fig.~{\ref{fig7}} and {\ref{fig8}}, we plot the variation of shadow radius $R_s$ and distortion parameter with the quintessence parameter $\omega_q$ for different values of normalization factor $\gamma$. The shadow radius $R_s$ increases with the $\omega_q$ while it decreases for the increasing  $\gamma$ that we have shown in the  Fig.~{\ref{fig7}}. The distortion parameter increases with $\omega_q$ (cf.~Fig.~{\ref{fig8}}). The black hole shadow appears more distorted in the presence of quintessential field. 

\begin{figure}
 \includegraphics[scale=0.65]{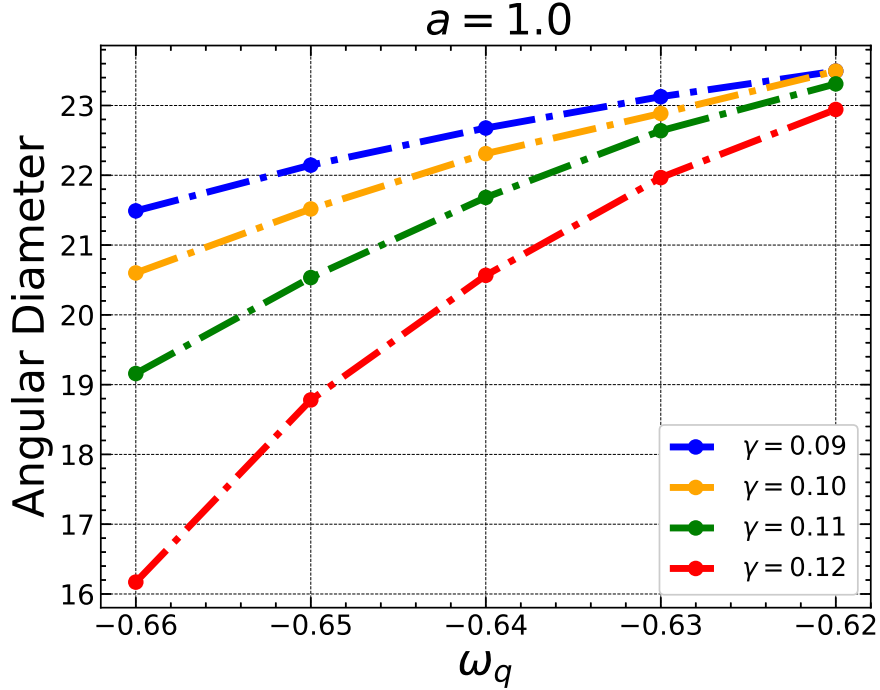}
  \caption{\label{fig10} Plot showing the variation of angular diameter of the black hole shadow  with the quintessential dark energy parameter $\omega_q$. }
 \end{figure}

The angular diameter $\theta_{d}$ of the black hole shadow can be estimated via
\begin{equation}
    \theta_d=\frac{2}{r_o}\times \sqrt{\frac{A}{\pi}},
    \label{angular}
\end{equation}
where $A$ is the area of the black hole shadow in quintessential dark energy. The variation of area of the black hole shadow $A$ with the quintessence parameter $\omega_q$ have shown in the Fig.~(\ref{fig9}) and Table~\ref{table4}. The  shadow area $A$ monotonically increase with the parameter $\omega_{q}$ while it decreases with the parameter $\gamma$ (cf.~Fig.~\ref{fig9} and Table~\ref{table4}). Using the Eq.~(\ref{angular}), we analytically estimate the angular diameter of the black hole shadow in quintessential dark energy. The angular diameter of the black hole shadow varies from $\theta_d \approx 21^{o}$ to $23^{o}$ with the  variation of $\omega_q$ from $-0.66$ to $-0.62$ for the parameter $\gamma=0.09$. Our results shows that the angular diameter of the black hole shadow monotonically increases with the quintessence dark energy parameter $\omega_q$. As we further increase the parameter $\gamma$ from $0.09$ to $0.12$, the angular diameter of the black hole shadow decreases, for $\gamma=0.12$ it varies from  $\theta_d  \approx 16^{o}$ to $22^{o} $ as clearly shown in the Fig.~{\ref{fig10}} and Table~\ref{table4}.

\section{Conclusion}\label{secfive}
The observations from EHT show that the angular size of the M87$^{*}$ black hole shadow is $42 \pm 3 \mu a s$, and   very recent results confirm the size of the supermassive black hole Sgr$^{*}$ $48.7 \pm 7  \mu a s$. These astrophysical objects are very far from the earth so the observed size of these black hole's shadow are very small such as 16.4 $megaparsec$ for M87 and 7.86 $kiloparsec$ for Sgr A${^*}$. Here,  we have analytically estimated the angular size of the quintessential dark energy black holes in the domain of outer communication. First, we derive the null geodesic equations of motion and obtained the effective potential for the black hole. After maximizing the effective potential we find impact parameters. We consider our observer in between the outer and cosmological horizon so we introduced corresponding  celestial coordinates and plotted various figures of black hole shadow for different values of quintessential dark energy parameter $\omega_q$, normalization factor $\gamma$ and spin parameter $a$. For the analytical study of the shape and size of the black hole shadow image, we numerically studied the shadow observables. In our study, we have found the size of the black hole shadow increases with the quintessence dark energy parameter $\omega_q$ while it decreases with the normalization factor $\gamma$. Next, the shape of the shadow, which is observable by the distortion parameter $\delta_s$, increases with the parameter $\omega_q$ while it decreases with the parameter $\gamma$. The observable shadow area $A$ also corresponds to the same results as the shadow radius and using the numerical results of the shadow area, we estimated the angular diameter of the quintessence dark energy black holes in the domain of outer communication. The angular diameter of the black hole shadow increases with the quintessence parameter $\omega_q$ and varies from $\theta_d \approx 16{^o}$ to 23$^o$ for $-0.66 \leq \omega_q \leq -0.62$ and $0.09 \leq \gamma \leq 0.12$ with spin parameter $a=1.0$ when the observer stays at rest in between the outer and the cosmological horizon of the quintessential dark energy black holes.

\section{Acknowledgement}
The author would like to thank Rahul Kumar Walia for the fruitful discussion and Tula's Institute for providing research facilities.

\end{document}